\documentclass[12pt, letterpaper]{article}
\usepackage{setspace}
\usepackage{caption}
\usepackage{color}
\usepackage{url}
\usepackage{verbatim}
\usepackage{subcaption}
\usepackage{amsmath,amssymb}
\usepackage{anyfontsize}
\usepackage{float}
\usepackage{listings}
\usepackage{multirow}
\usepackage{graphicx}

\title{HQAlign: Aligning nanopore reads for SV detection using current-level modeling}
\author{Dhaivat Joshi, Suhas Diggavi\footnote{SD is corresponding author: suhas@ee.ucla.edu}, Mark J.P. Chaisson, and Sreeram Kannan
\footnote{DJ and SD are at the University of California, Los Angeles.
MC is at the Department of Quantitative and Computational Biology, University of Southern California, Los Angeles.
SK is at the University of Washington, Seattle.}}
\date{}

\begin{document}

\maketitle

\begin{abstract}

\noindent\textbf{Motivation:}
Detection of structural variants (SV) from the alignment of sample DNA reads to the reference genome is an important problem in understanding human diseases. Long reads that can span repeat regions, along with an accurate alignment of these long reads play an important role in identifying novel SVs. Long read sequencers such as nanopore sequencing can address this problem by providing very long reads but with high error rates, making accurate alignment challenging. Many errors induced by nanopore sequencing have a bias because of the physics of the sequencing process and proper utilization of these error characteristics can play an important role in designing a robust aligner for SV detection problems. In this paper, we design and evaluate HQAlign, an aligner for SV detection using nanopore sequenced reads. The key ideas of HQAlign include (i) using basecalled nanopore reads along with the nanopore physics to improve alignments for SVs (ii) incorporating SV specific changes to the alignment pipeline (iii) adapting these into existing state-of-the-art long read aligner pipeline, minimap2 (v2.24), for efficient alignments.\\
\textbf{Results:}
We show that HQAlign captures about $4-6\%$ complementary SVs across different datasets which are missed by minimap2 alignments while having a standalone performance at par with minimap2 for real nanopore reads data.
For the common SV calls between HQAlign and minimap2, HQAlign improves the start and the end breakpoint accuracy for about $10-50\%$ of SVs across different datasets.
Moreover, HQAlign improves the alignment rate to $89.35\%$ from minimap2 $85.64\%$ for nanopore reads alignment to recent telomere-to-telomere CHM13 assembly, and it improves to $86.65\%$ from $83.48\%$ for nanopore reads alignment to GRCh37 human genome.\\
\textbf{Availability:} \url{https://github.com/joshidhaivat/HQAlign.git}

\end{abstract}
\section{Introduction}

Structural variations (SVs) are genomic alterations of size at least 50 bp long, including insertions, deletions, inversions, duplications, translocations or a combination of these types \cite{alkan2011genome}. The study of these genetic variations has an important role in understanding human diseases, including cancer \cite{icgc2020pan}, and begins with sequence alignment from the sample back to the reference genome. Accurate alignment of short reads from high throughput sequencing poses a challenge, especially, in the repetitive regions of the genome which are also the hotspots of nearly $70\%$ of the observed structural variations \cite{rowell2019comprehensive}.

Long read sequencing technologies have addressed this problem by producing reads that are longer than the repeat regions, therefore, enabling the detection of variants in the repeat regions at the cost of higher error rates than short read sequencing technologies. This high error rates in the long reads lead to non-contiguous alignment which poses a challenge in variant detection problem, especially, in the repeat regions.

\begin{figure*}[!t]
	\centering
	\begin{minipage}{0.7\textwidth}
		\centering
		\includegraphics[width=\linewidth]{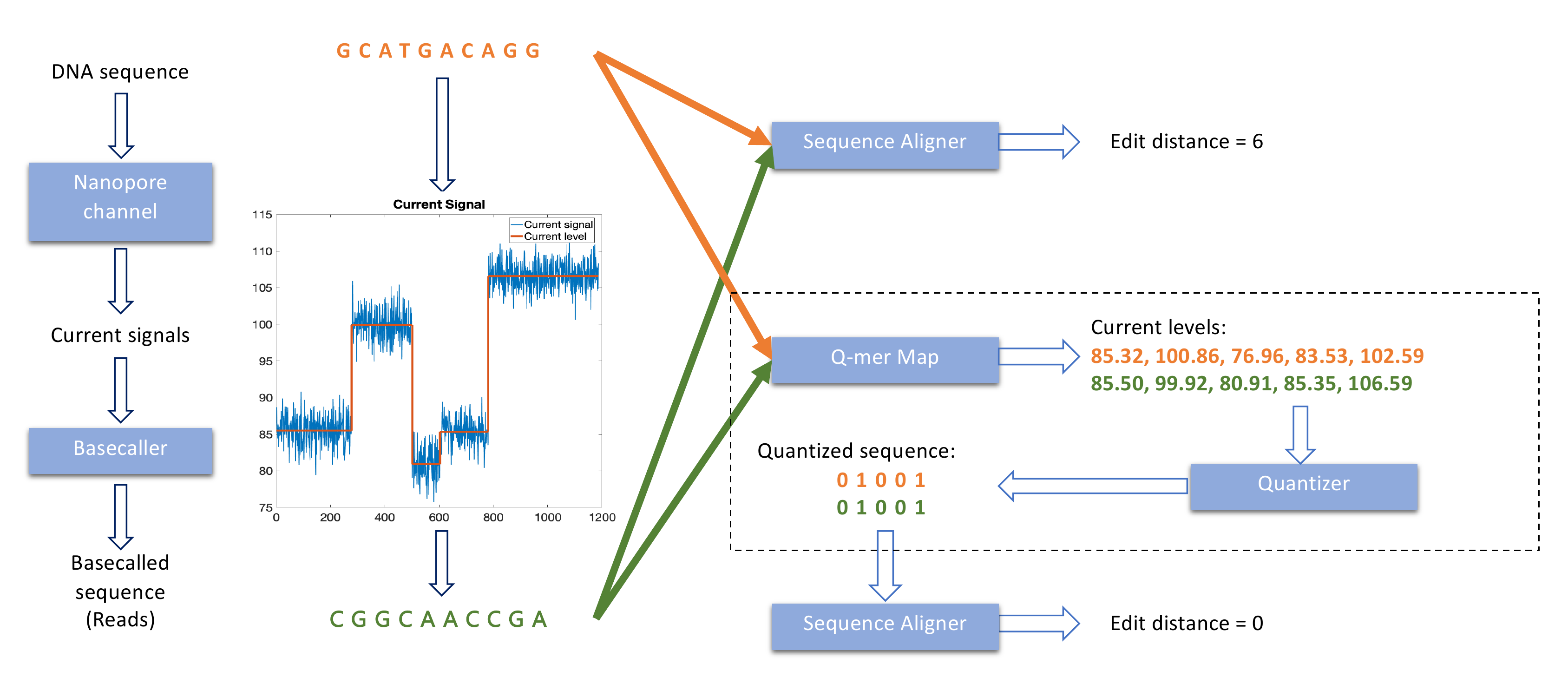}
		\centering
		(a)
	\end{minipage}\hfill
	\begin{minipage}{0.29\textwidth}
		\centering
		\includegraphics[width=\linewidth]{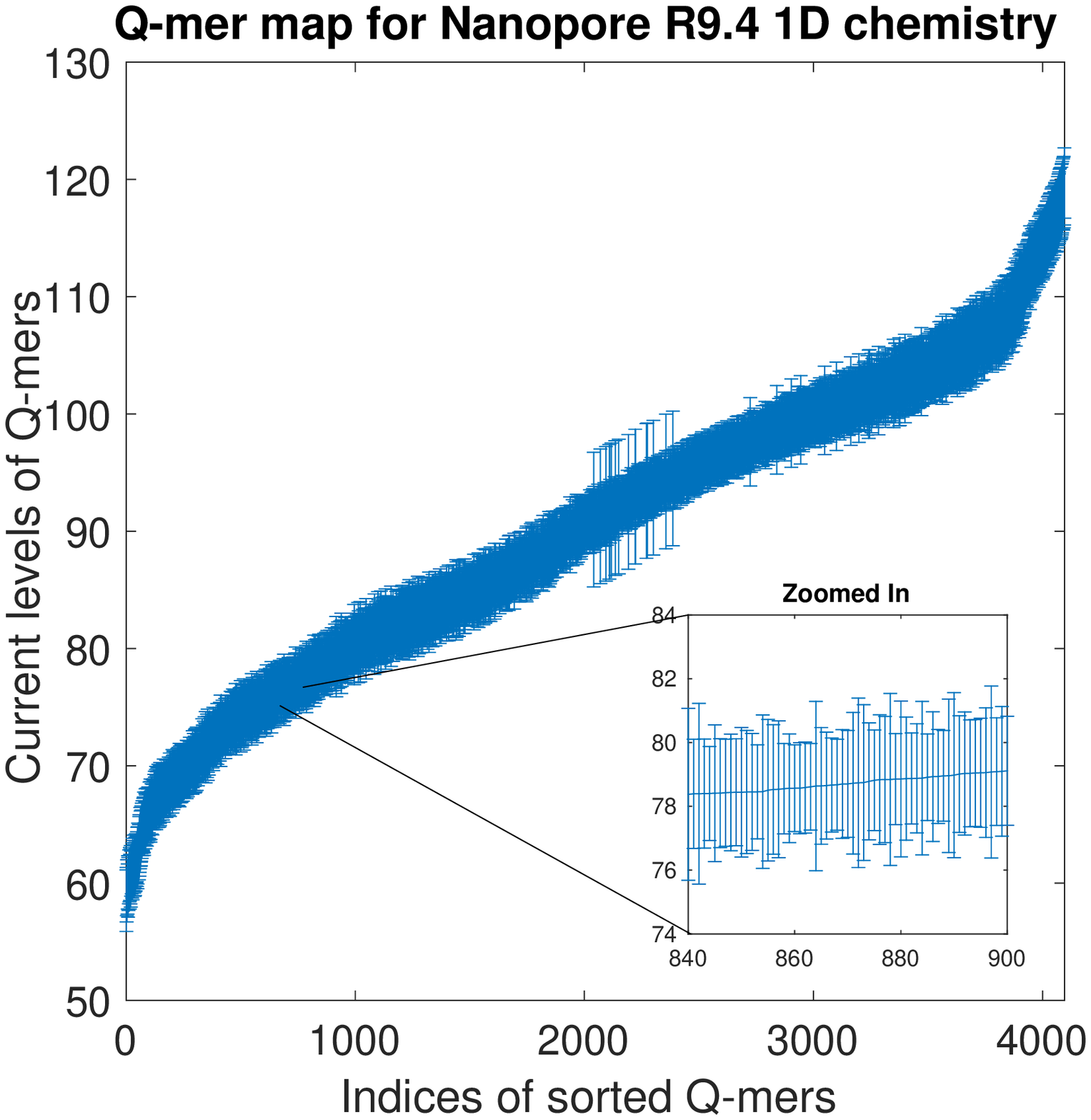}
		\centering
		(b)
	\end{minipage}\hfill
	\begin{minipage}{0.99\textwidth}
		\includegraphics[width=\linewidth]{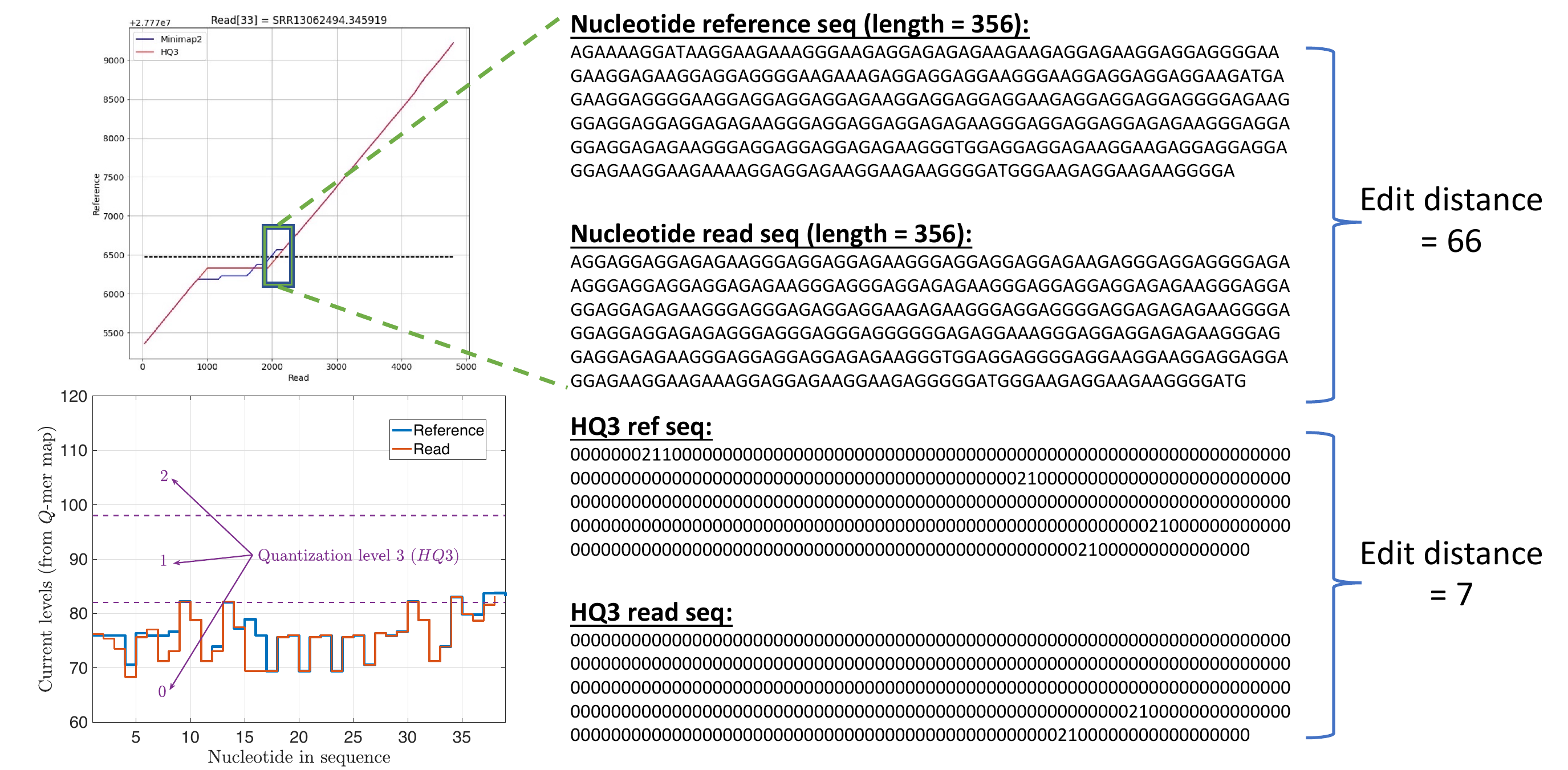}
		\centering
		(c)
	\end{minipage}\hfill
	\caption{(a) An example to illustrate the error biases in nanopore basecalled reads which can be resolved through the $Q$-mer map ability of HQAlign to perform accurate alignment despite of the errors (the edit distance used here is domain specific and is used to demonstrate accuracy of the alignment). (b) $Q$-mer map for Nanopore R9.4 1D flow cell (for $Q=6$). It represents the physics of nanopore. The median current value along with the standard deviation (as error bars) are plotted for all $6$-mers in the $Q$-mer map for R9.4 1D nanopore flow cell (the $Q$-mers are sorted in increasing median current levels). Note that the difference between the median current levels of any two consecutive $Q$-mers is very small, therefore, resulting in large overlaps. (c) An example from PromethION R9.4.1 ONT data in the neighborhood of a SV in repeat region showing the two different nucleotide sequences have similar current levels and therefore, the edit distance as observed through the lens of quantized sequence is significantly lower in $HQ3$.}
	\label{fig:introduction}
\end{figure*}

Nanopore sequencing \cite{Mikheyev2014,Deamer2016} is a long read sequencing technology that provides reads (with average read length $~10$-kb and the longest read sequenced more than $2$-Mb long) that can span these repetitive regions but it has a high error rate of (average) $10\%$.
This high error rates result in low accuracy alignments \cite{Krianovi2017} using state-of-the-art methods including minimap2 (v2.24) \cite{li2021new} which is a fast method designed for the computationally challenging task of long sequence alignment. This problem is further amplified in the repetitive regions such as variable-number tandem repeats (VNTR) region that accounts for a significant fraction of SVs \cite{chaisson2019multi,ebert2021haplotype}.
However, these errors in nanopore sequencing have a bias induced from nanopore physics which is missed by many long read aligners since they consider the errors as independent insertions, deletions, and substitutions. In nanopore sequencing, a DNA strand migrates through the nanopore, and an ionic current according to the nucleotide sequence in or near the nanopore is established. However, because of the physics and non-idealities of the nanopore sequencing, each current level recorded depends on a $Q$-mer (a set of $Q$ consecutive nucleotide bases which influence the measurement in the nanopore) \cite{Laszlo2014,NanoporeHowItWorks,MaoDigKanIT18}. These current readings are translated back to nucleotide sequences by basecalling algorithms. Therefore, the error biases could be introduced in basecalling, especially, between different $Q$-mers that have similar current levels. This similarity in the median current levels for different $Q$-mers is captured by the $Q$-mer map as shown in Figure \ref{fig:introduction}b. A $Q$-mer map represents the median current level for different $Q$-mers ($Q=6$) for nanopore flow cell. It is evident from this figure that there is a significant overlap between the current levels observed for different $Q$-mers migrating the nanopore. We propose a new alignment method, HQAlign (which is based on QAlign \cite{joshi2021qalign}), which is designed specifically for detecting SVs while incorporating the error biases inherent in the nanopore sequencing process. HQAlign pipeline is modified to enable detection of inversion variants which was not feasible with the earlier QAlign pipeline (refer to methods section 2.2 for details).

HQAlign takes the dependence of $Q$-mer map into account to perform accurate alignment with modifications specifically for discovery of SVs. Figure \ref{fig:introduction}a gives an example where a DNA sequence (GCATGACAGG) is sequenced incorrectly as (CGGCAACCGA) due to the error bias in nanopore sequencer. Therefore, the sequences are different in the nucleotide space but they are identical in the $Q$-mer map space. It is important to note that \emph{no additional soft information is used} to establish this identity such as raw nanopore current values for the nanopore reads. Instead, the nucleotide sequences that have indistinguishable current levels from the lens of the $Q$-mer map are mapped to a common quantized sequence. A nucleotide sequence is converted to a quantized sequence by first converting the nucleotide sequence to a sequence of current levels using the $Q$-mer map and then converting the sequence of (continuous) current levels to a (finite level) quantized sequence by hard thresholding the current levels (refer to Supplementary material section A.1 for more details). Therefore, the additional information about the raw current signals is not used in the quantization process but only the $Q$-mer map is utilized. This process is explained in detail in Supplementary material Figure \ref{fig:quantization}.
Further, the quantization of continuous current levels to finite discrete levels enables the use of existing software pipelines of state-of-the-art long read aligners such as minimap2 as the core seed and extend algorithm for the alignment of quantized sequences.

In HQAlign, we first perform the alignment of reads onto genome using minimap2 to determine the region of interest where a read can possibly align to, and then re-align the quantized read to the quantized genome region from the first step. This helps in performing an accurate alignment of the read to the region of genome without dropping the frequently occurring seed matches from the chain in minimap2 algorithm while taking the error biases of nanopore sequencing into account through quantized sequences.
Moreover, HQAlign pipeline enables detection of inversion variants unlike QAlign pipeline. In QAlign, the quantized reverse complement of read is aligned separately to the quantized genome, therefore, the alignment of inverted sequence is not observed in QAlign (as shown in Figure \ref{fig:methods}a). However, in HQAlign, we have modified the minimap2 pipeline to align the reverse complement of quantized read along with the aligning the forward quantized read sequence to the quantized genome, simultaneously. This is necessary for detection of inverted alignment using quantized sequences because unlike nucleotide sequences where minimap2 can inherently produce the reverse complement of the input nucleotide query, the reverse complement of the quantized sequence is to be computed separately.
Further, HQAlign is about 2x faster than QAlign as the seed search domain is reduced to a region of interest determined in the first step of the pipeline for the quantized sequences (as shown in Table \ref{table:computation_time}).

Figure \ref{fig:introduction}c demonstrate an example from real ONT reads data in a repeat region (note that a pattern of a few consecutive nucleotide bases is repeated in the example) that is flanking around an insertion structural variant. Minimap2 alignment of nucleotide reference and read (both of length 356 from the region highlighted with a box) have an edit distance of $66$ whereas the $HQ3$ alignment ($HQ3$ is an alignment from HQAlign pipeline where the nucleotide sequences are translated to three level quantized sequences, refer to section 2.2 for details) of quantized reference and read sequences from the same region have a significantly smaller edit distance of $7$. This is because the current level sequence (by converting the nucleotide sequences using the $Q$-mer map in Figure \ref{fig:introduction}b) for the reference and the read are very similar. Therefore, the sequences that are far apart in nucleotide space are inherently very similar in the $HQ3$ space in terms of the edit distance in the transformed space.

We show that HQAlign gives significant performance improvements in quality of read alignment across real and simulated data. The well-aligned reads (a read is defined as well aligned if at least $90\%$ of the read is aligned on the genome with a mapping quality more than $20$) improves to $86.65\%$ with $HQ3$ from $83.48\%$ with minimap2 (v2.24) for the alignment of ONT reads from HG002 sample to GRCh37 human genome. The metric improves to $89.35\%$ from $85.64\%$ for HG002 reads alignment to T2T CHM13 assembly \cite{rhie2022complete}, and improves to $81.57\%$ from $81.01\%$ for the simulated reads data. These results are presented in the results section 3.1.2 Table \ref{table:alignment}.

In terms of SV detection, HQAlign has F1 score at par with minimap2 (v2.24) with Sniffles2 \cite{smolka2022comprehensive} as the variant calling algorithm across both real and simulated dataset (Table \ref{table:sv_precision_recall}). However, both HQAlign and minimap2 captures many complementary calls ($~4-6\%$) which are missed by the other method (as shown in Figure \ref{fig:complimentary_chm13}, \ref{fig:complimentary_giab}, \ref{fig:complimentary_hg002}, \ref{fig:complimentary_sim}, \ref{fig:complimentary_chm13-cen}). For instance, the complementary HQAlign calls are SVs that are uniquely called by HQAlign or labeled missed in minimap2 due to breaking in the SV and vice-versa for the complementary calls in minimap2. Further, the analysis of common true positive SV calls in HQAlign and minimap2 against the truth set shows that HQAlign has on average a significant improvement ($10-50\%$, from the slope of the regression line in Figures \ref{fig:sv_chm13}, \ref{fig:sv_giab}, \ref{fig:sv_hg002}, \ref{fig:sv_chm13-centromere}, and weighted average across all datasets for $~39\%$ SVs)
in the breakpoint accuracy than minimap2 for the calls with difference in breakpoint greater than 50 bp (breakpoint accuracy is determined from the difference in the start and end breakpoints of a SV with respect to the match SV in truth set, therefore, lower the difference higher is the breakpoint accuracy, refer to section 2.3 for precise definition). Moreover, for the common true positive calls, HQAlign has (on average) better SV length similarity than minimap2 (when SV length similarity is less than $0.95$, SV length similarity is a measure of how similar is the length of SV from an alignment method relative to the match SV in truth set; refer to section 2.3 for a precise definition) as shown in Figures \ref{fig:sv_chm13}, \ref{fig:sv_giab}, \ref{fig:sv_hg002}, and \ref{fig:sv_chm13-centromere}.

\vspace{-3mm}
\section{Methods}\label{section:Methods}

\begin{figure}[!h]
	\begin{minipage}{0.4\columnwidth}
		\centering
		\includegraphics[width=\linewidth]{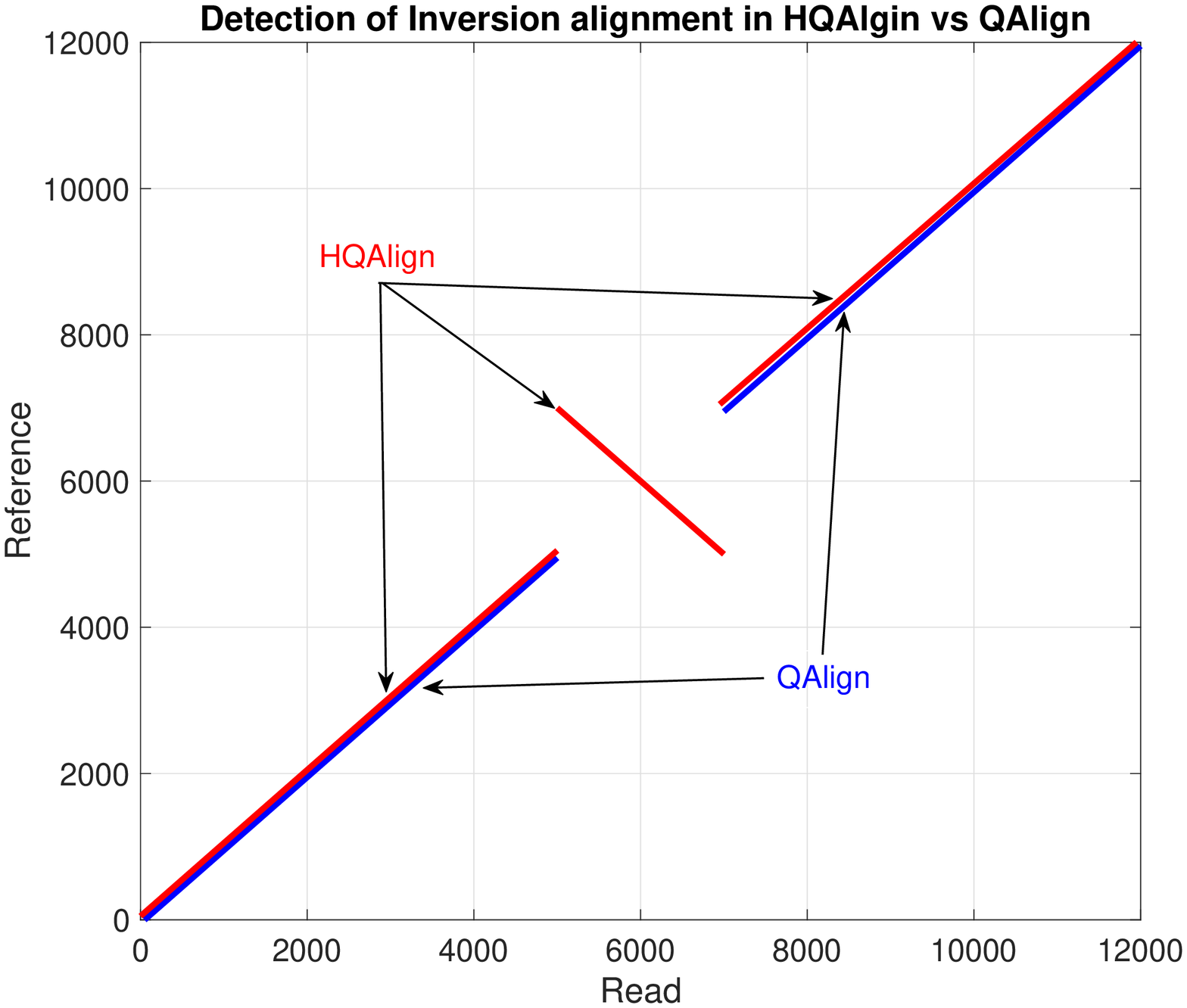}\\
		\centering
		(a)\vspace{1mm}
	\end{minipage}
	\begin{minipage}{0.6\columnwidth}
		\centering
		\includegraphics[width=\linewidth]{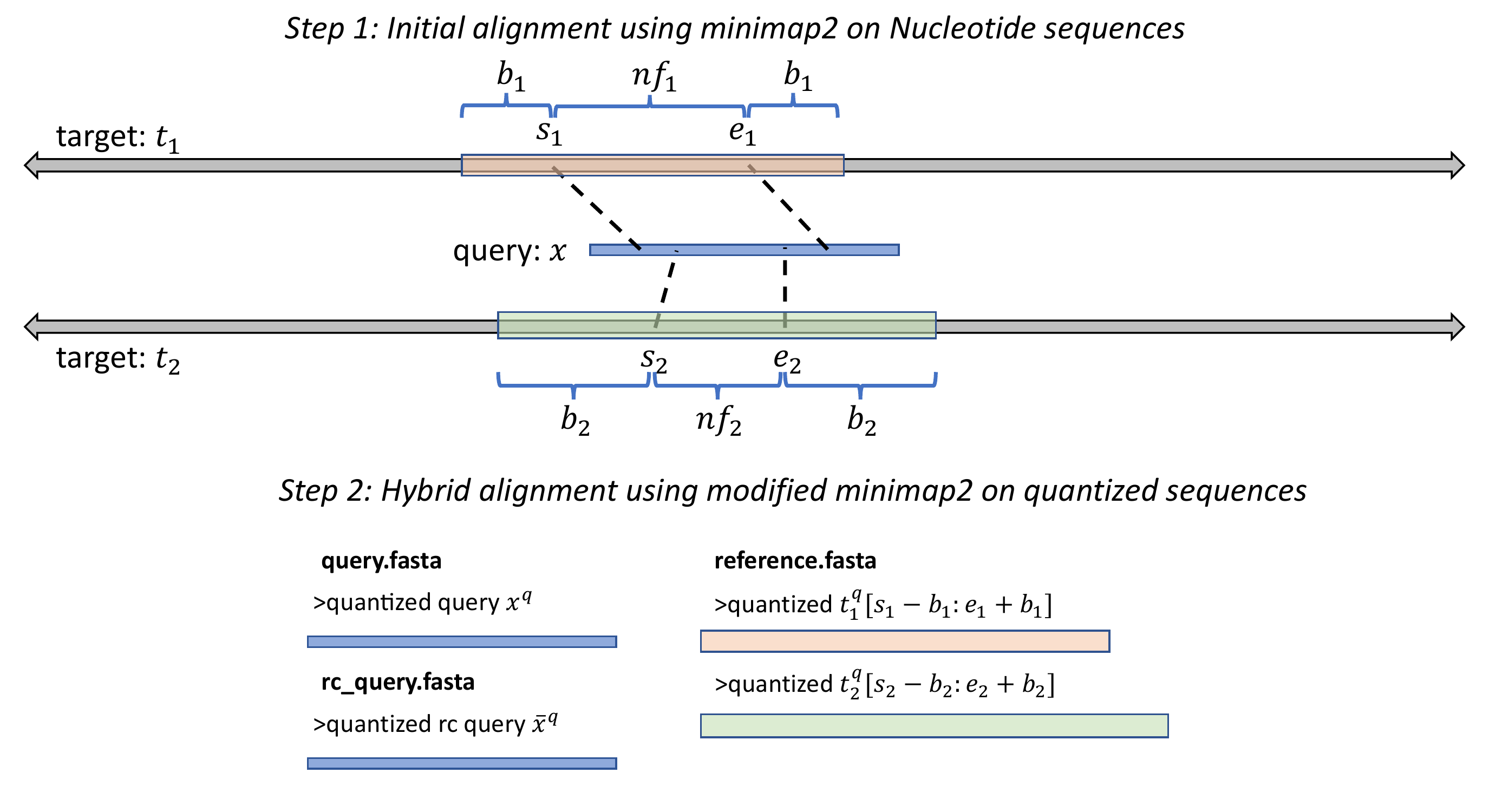}
		\centering
		(b)
	\end{minipage}\\\hfill
	\begin{minipage}{0.5\columnwidth}
		\centering
		\includegraphics[width=\linewidth]{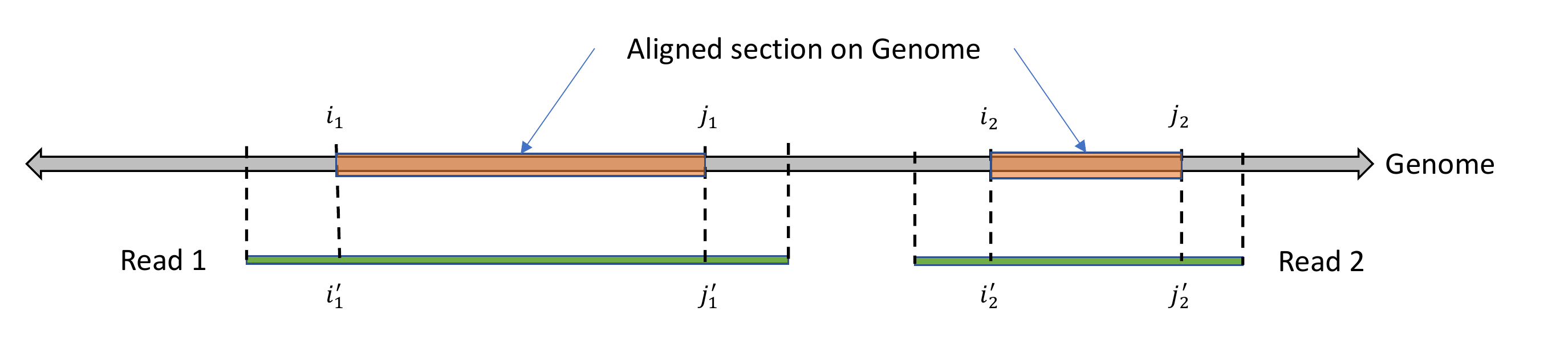}
		\centering
		(c)
	\end{minipage}
	\begin{minipage}{0.5\columnwidth}
		\centering
		\includegraphics[width=\linewidth]{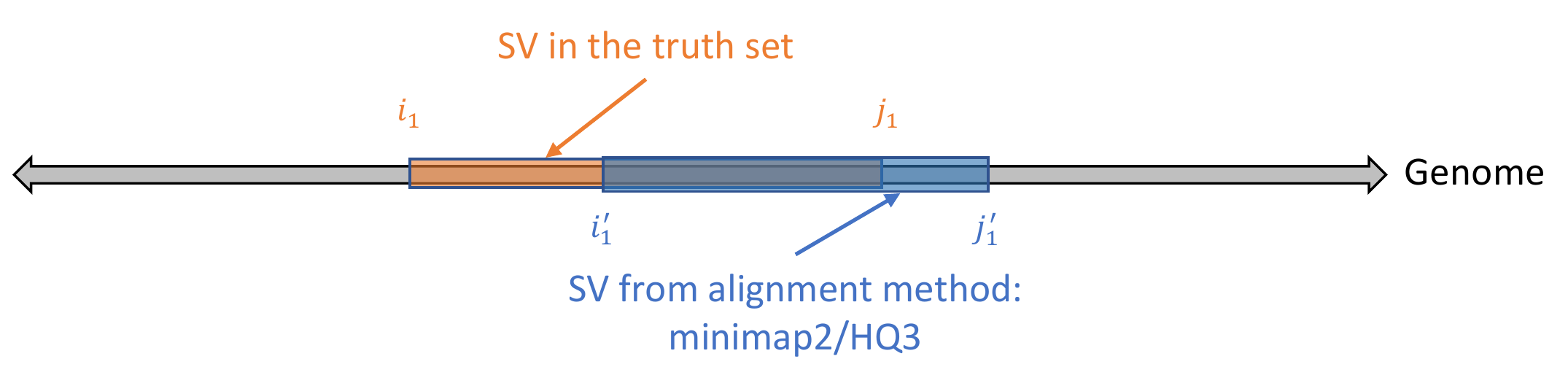}
		\centering
		(d)
	\end{minipage}\\\hfill
	\caption{(a) An example to demonstrate the ability of HQAlign pipeline to align inverted sequences where QAlign fails (b) An example of HQAlign pipeline. (c) An example of read-to-genome alignment. (d) Comparison of SV in truth set to SV determined by method: minimap2/$HQ3$.}
	\label{fig:methods}
\end{figure}

The HQAlign strategy consists of two steps: (1) the initial alignment of the standard basecalled query sequence $x$ to a target sequence $t$ using Minimap2. This primitive step identifies the region of interests on the target sequence where $x$ aligns. (2) the hybrid step is re-aligning the query $x$ only to the region of interest determined in the first step using a modified pipeline for QAlign method. QAlign method converts the nucleotide sequences in both query $x$ and target region $t$ to quantized (three levels) sequence, and then aligns these quantized sequences using any state-of-the-art aligner. We have modified the minimap2 (v2.24) pipeline for the alignment of the quantized sequences in the second step so that it enables detection of the inversion variants in alignment using the quantized sequences. This is explained in detail in section 2.2. Further, this strategy is about 2x faster than QAlign standalone as it narrows down the seed search domain for lower alphabet size (\emph{e.g.} three levels) in QAlign. This strategy is explained in Figure \ref{fig:methods}b, and mathematically in the following sections.

\begin{table*}[!h]
\centering
\caption{Comparison of computation time for alignment of 500k randomly sampled ONT reads to CHM13 assembly using 20 threads for each method.}
\medskip
\begin{tabular}{|p{4.5cm}|p{2.8cm}|p{2.8cm}|p{2.8cm}|}
\hline
\bf{Method} & \bf{minimizer (seed) length} &\bf{Real time (in seconds)} & \bf{CPU time (in seconds)}\\
\hline
%
  %
\multirow{1}{=}{QAlign ($Q3$)}
  & 18
  & $10,312$
  & $177,516$
  \\\hline
\multirow{1}{=}{HQAlign ($HQ3$)}
  & 18
  & $5,033$
  & $76,375$
  \\\hline
\end{tabular}
\label{table:computation_time}
\end{table*}

\subsection{Initial alignment}
The nucleotide query $x$ is aligned to a nucleotide target sequence $t$ using minimap2. This is similar to aligning a read to a genome with one chromosome. Here we consider only one chromosome in target $t$ for simplicity but the method generalizes to multiple chromosomes in $t$ such as $t = (t_1,t_2,\dots,t_m)$ (this generalization is explained in detail in Supplementary material section A.2). This step identifies the region of interests on the target $t$, say, $t[s_i:e_i]$, where $i \in \{1,2,3,\dots\}$ represent one or more alignments on $t$ and $s_i$ and $e_i$ are the corresponding start and end location of each alignment $i$ on target $t$, respectively.

\subsection{Hybrid alignment}
In this step, the query $x$ is re-aligned to an extended region of interest on the target $t[s_i^q:e_i^q]$ using the modified pipeline for QAlign method, where $s_i^q = s_i - b_i$ and $e_i^q = e_i + b_i$, $b_i = (1-f_i+0.25)n$ is an appended extension of the region of interest on target, $f_i = (e_i-s_i)/n$ is the fraction of read aligned in initial step, and $n$ is the length of the query $x$. The nucleotide query $x$ and the nucleotide extended target $t[s_i^q:e_i^q]$ are converted to the quantized query $x^q$ and quantized extended target $t^q[s_i^q:e_i^q]$, respectively, using the quantization method demonstrated in QAlign (refer to Supplementary material section A.1 for more details on quantization process).
It is important to note that we do not use any additional soft information such as raw current signals from nanopore sequencing in the quantization process, instead, we translate the basecalled nucleotide reads to current levels using the $Q$-mer map (in Figure \ref{fig:introduction}b) and then hard threshold the current levels to finite (\emph{e.g.} three) levels to get the quantized ($HQ3$) reads (refer Supplementary material Figure \ref{fig:quantization} for more details on quantization process). These quantized sequences are then aligned using a modified pipeline of minimap2 (v2.24). We have modified minimap2 pipeline for this hybrid step to accept the quantized reverse complement query $\bar x^q$ as an input which helps in indentifing the inversion SVs in contiguous alignment with quantized sequences which was not possible with the earlier QAlign method as shown in Figure \ref{fig:methods}a. QAlign uses the default minimap2 pipeline for the alignment of quantized sequences which inherently aligns both the forward and the reverse complement strand of the sequences in nucleotide domain. However, the quantized reverse complement sequence cannot be computed given only the forward quantized sequence, therefore, QAlign separately aligns both quantized forward and quantized reverse complement sequence. This method, however, fails to identify an inverted alignment as shown in Figure \ref{fig:methods}a. Therefore, in HQAlign, we have modified the minimap2 pipeline to enable alignment using both quantized forward and quantized reverse complement sequence, simultaneously.
Note that the quantized alignment employs a different minimizer length $\texttt{k}=18$ in minimap2 for ternary ($HQ3$) quantization.

We define several metrics that are used for the performance evaluation of HQAlign against minimap2 (these metrics are used from the earlier QAlign method \cite{joshi2021qalign}).
\begin{enumerate}
\item[(i)] \textbf{well-aligned:} Consider in Figure \ref{fig:methods}c, \textit{Read 1} aligns at location $i_1$ through $j_1$ on the genome determined using nucleotide alignment. We say that the read is well-aligned, if at-least $90\%$ of the read is aligned onto the genome (i.e., $j_1{-}i_1 {\geq} 0.9(\text{length}(\textit{Read 1}))$), and has high mapping quality (greater than $20$). This metric quantifies the reads that are mapped almost entirely to the reference.

\item[(ii)] \textbf{normalized edit distance:} In order to compare the quality of the alignments at fine-grained level, we further define normalized edit distance.
The normalized edit distance for nucleotide alignment is defined as
\begin{equation}
{\small\frac{\text{edit\_distance}\{ r; G[i_1:j_1]\}}{\text{length}(r)}}
\end{equation}
and for quantized alignment is
\begin{equation}
{\small\frac{\text{edit\_distance}\{ r; G[i_{1}^{q}:j_{1}^{q}]\}}{\text{length}(r)}}
\end{equation} 
where $i_1, j_1$ are the start and end location of alignment on genome in nucleotide space and $i_1^q, j_1^q$ are the start and end location of alignment on genome in the quantized space, $r$ is the entire read and $G$ is the genome as shown in Figure \ref{fig:methods}c.
It is important to note that for computing the normalized edit distance for alignments in the quantized space, we only leverage the information of location of the alignment on genome from quantized space, \textit{i.e.} $i_1^q$ and $j_1^q$, but the edit distance between read and the aligned section on genome is computed on the nucleotide sequences. This metric gives a measure of the distance similarity between two sequences, especially, used for the real data where the truth of sequence sampling location is not known.

\item[(iii)] \textbf{normalized alignment length:} Another metric at the fine-grained level is normalized alignment length, which is the ratio of the length of the section on genome where a read aligns to the length of the read. It is
\begin{equation}
{\small\frac{j_1-i_1}{\text{len}(r)}}
\end{equation}
for nucleotide alignment, and
\begin{equation}
{\small\frac{j_1^q-i_1^q}{\text{len}(r^Q)}}
\end{equation}
for quantized alignment. A contiguous alignment tends to have this metric as $1$. This metric gives a measure of the contiguity of the alignment.
\end{enumerate}

\subsection{SV calling}
The alignments from HQAlign and minimap2 in sorted \textit{bam} format are used to detect structural variants using Sniffles2. These calls are benchmarked against a truth set using Truvari \cite{english2022truvari}. We have used F1 score, precision and recall as the metric to analyze the performance of HQAlign and compare them with minimap2. Precision ($P$) is defined as the fraction of SVs detected by the algorithm in the truth set among the total SVs detected by the algorithm. Recall ($R$) is the fraction of SVs  detected by the algorithm in the truth set among the total SVs in the truth set. F1 score is the harmonic mean of precision and recall ($=\frac{2P\cdot R}{P+R}$). Further, we have observed that there are many complementary SV calls made by both minimap2 and $HQ3$ that are missed by the other method. Therefore, we have defined a union model which takes a union of the SV calls from both minimap2 and $HQ3$. The precision, recall, and F1 score of the union model are also computed and reported in Table \ref{table:sv_precision_recall}.

Further, the quality of the SVs for the common calls in minimap2 and HQAlign is evaluated by comparing the following metrics w.r.t. the SVs in truth set
\begin{enumerate}
\item[(i)] \textbf{breakpoint accuracy:} Breakpoint accuracy is measured by taking an average of the difference in the in the start and end breakpoint of the SV w.r.t. the SV in truth set. For instance, as shown in Figure \ref{fig:methods}d, $i_1$ and $j_1$ are the start and the end point on genome of SV in the truth set, and $i_1^{'}$ and $j_1^{'}$ are the start and the end point of the same SV determined by any alignment method (minimap2/$HQ3$), then breakpoint score is calculated as
\begin{equation}
{\small\frac{|i_1^{'}-i_1|+|j_1^{'}-j_1|}{2}}
\end{equation}
where $|\cdot|$ is absolute value function. Therefore, lower the score higher is the breakpoint accuracy of the SV determined by the alignment method.

\item[(ii)] \textbf{SV length similarity:} SV length similarity is measured as the ratio of minimum SV length in truth set and from algorithm to the maximum of two values. Mathematically, it is
\begin{equation}
{\small\frac{\min(j_1-i_1,j_1^{'}-i_1^{'})}{\max(j_1-i_1,j_1^{'}-i_1^{'})}}
\end{equation}
for the example shown in Figure \ref{fig:methods}d.

\end{enumerate}

\section{Results}

In this section, we demonstrate the results for (1) comparison of alignments from HQ3 and minimap2 on real as well as simulated data, and (2) comparison of SV calls from HQ3 and minimap2 alignments using Sniffles2 as the variant caller on real and simulated data.

\subsection{DNA read-to-genome alignment}

\subsubsection{Datasets}
We have used the publicly available R9.4.1 ONT PromethION reads dataset from HG002 sample \cite{ren2021lra}. These reads are aligned to the recent telomere-to-telomere assembly CHM13 and to the human reference genome GRCh37. GRCh37 is used as the reference build to map the real data so that the curated variants can be used for accuracy analysis \cite{zook2020robust}. Further, we have also benchmarked the performance of HQAlign and minimap2 on simulated data for both alignments and SV calling.

\subsubsection{Alignment results}

\begin{figure*}[!h]
	\begin{minipage}{0.5\linewidth}
		\centering
		\includegraphics[width=\linewidth]{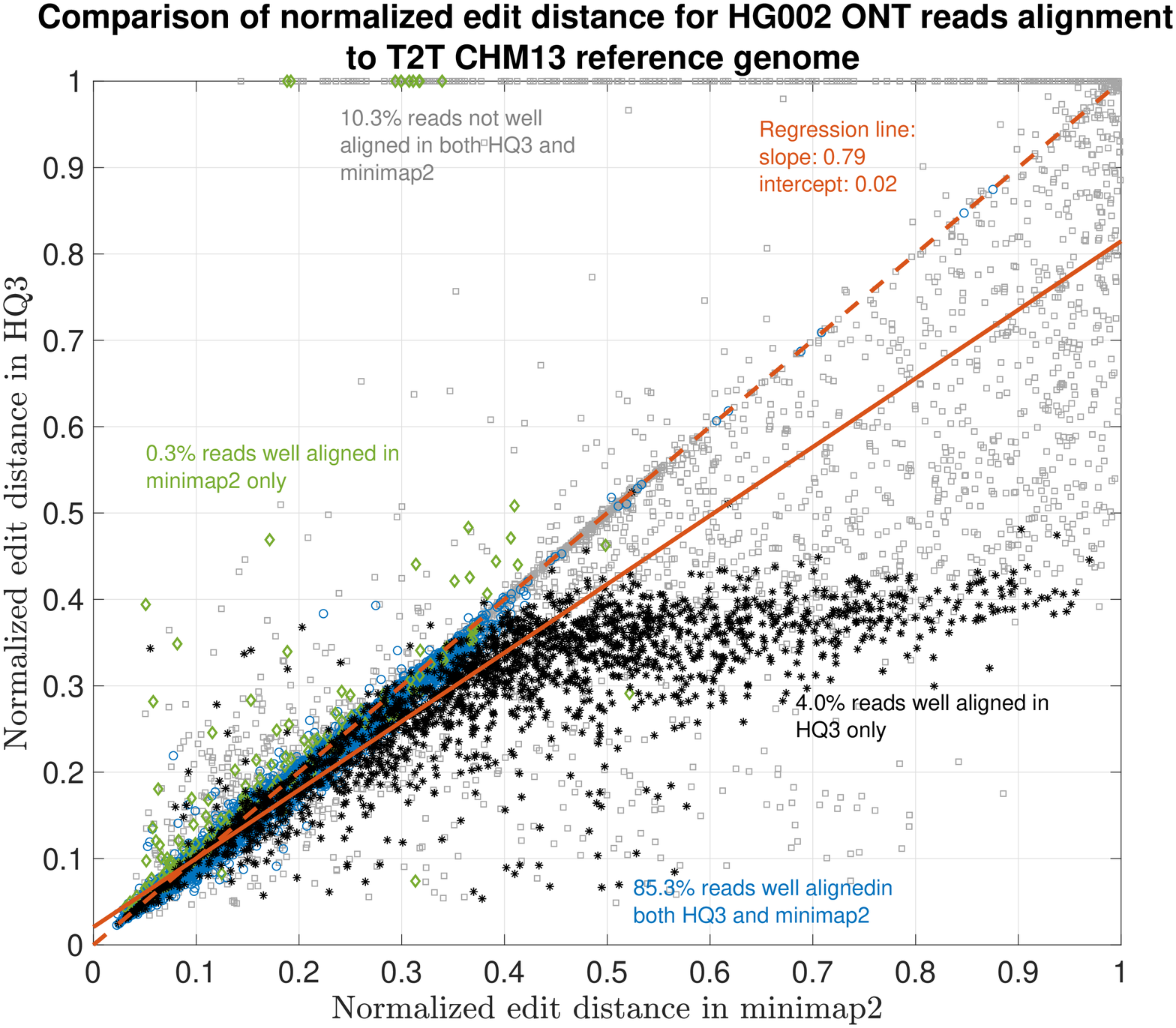}
		\label{fig:alignment_ned_chm13}
		(a)
	\end{minipage}\hfill
	\begin{minipage}{0.5\linewidth}
		\centering
		\includegraphics[width=\linewidth]{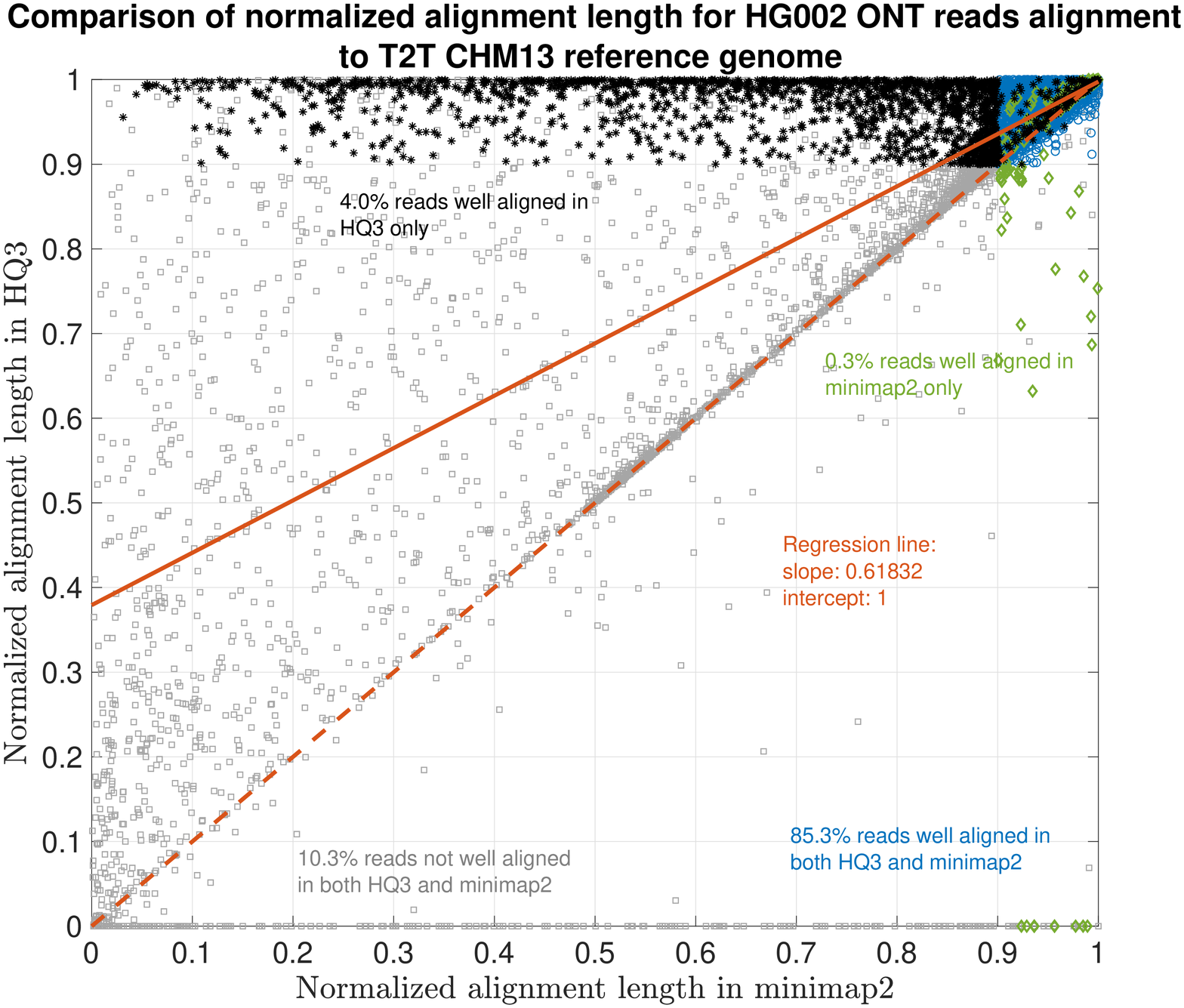}
		\label{fig:alignment_nal_chm13}
		(b)
	\end{minipage}\hfill
	\caption{{\bf HG002 nanopore long DNA reads alignment onto T2T CHM13 genome.}
	(a) Comparison of normalized edit distance for HG002 R9.4.1 PromethION reads data. Smaller values for normalized edit distance is desirable as it represents better alignment. The slope of the regression line is $0.79<1$, therefore, representing better alignments with $HQ3$ than minimap2 alignments for same reads on average.
	(b) Comparison of normalized alignment length for HG002 R9.4.1 PromethION reads data. Normalized alignment length of $1$ is desirable as it represents that entire read is aligned. The majority of the reads are above $y=x$ line representing longer alignment length in $HQ3$ than minimap2 alignment.}
	\label{fig:alignment_chm13}
\end{figure*}

\begin{figure*}[!]
	\begin{minipage}{0.5\linewidth}
		\centering
		\includegraphics[width=\linewidth]{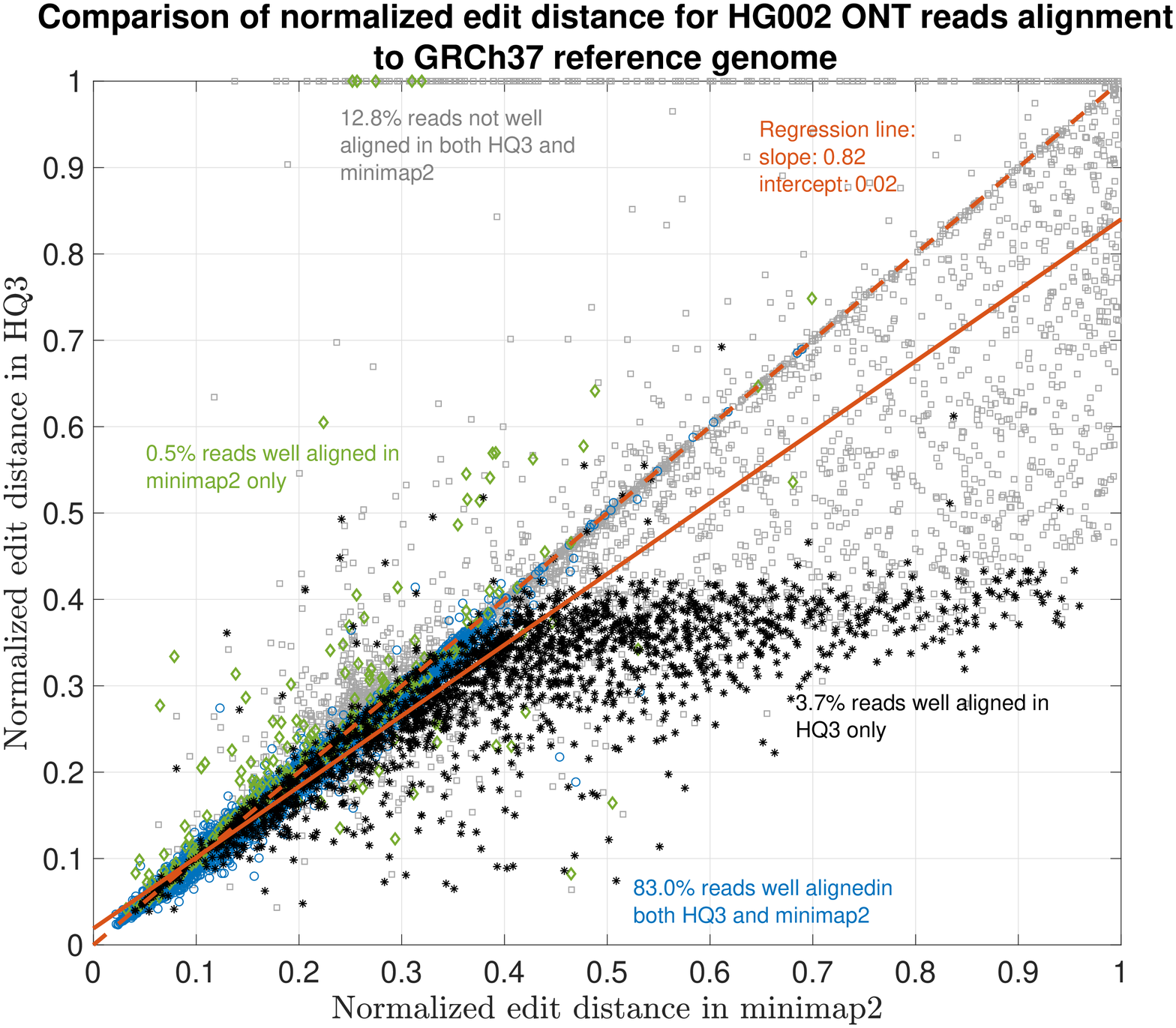}
		\label{fig:alignment_ned}
		(a)
	\end{minipage}\hfill
	\begin{minipage}{0.5\linewidth}
		\centering
		\includegraphics[width=\linewidth]{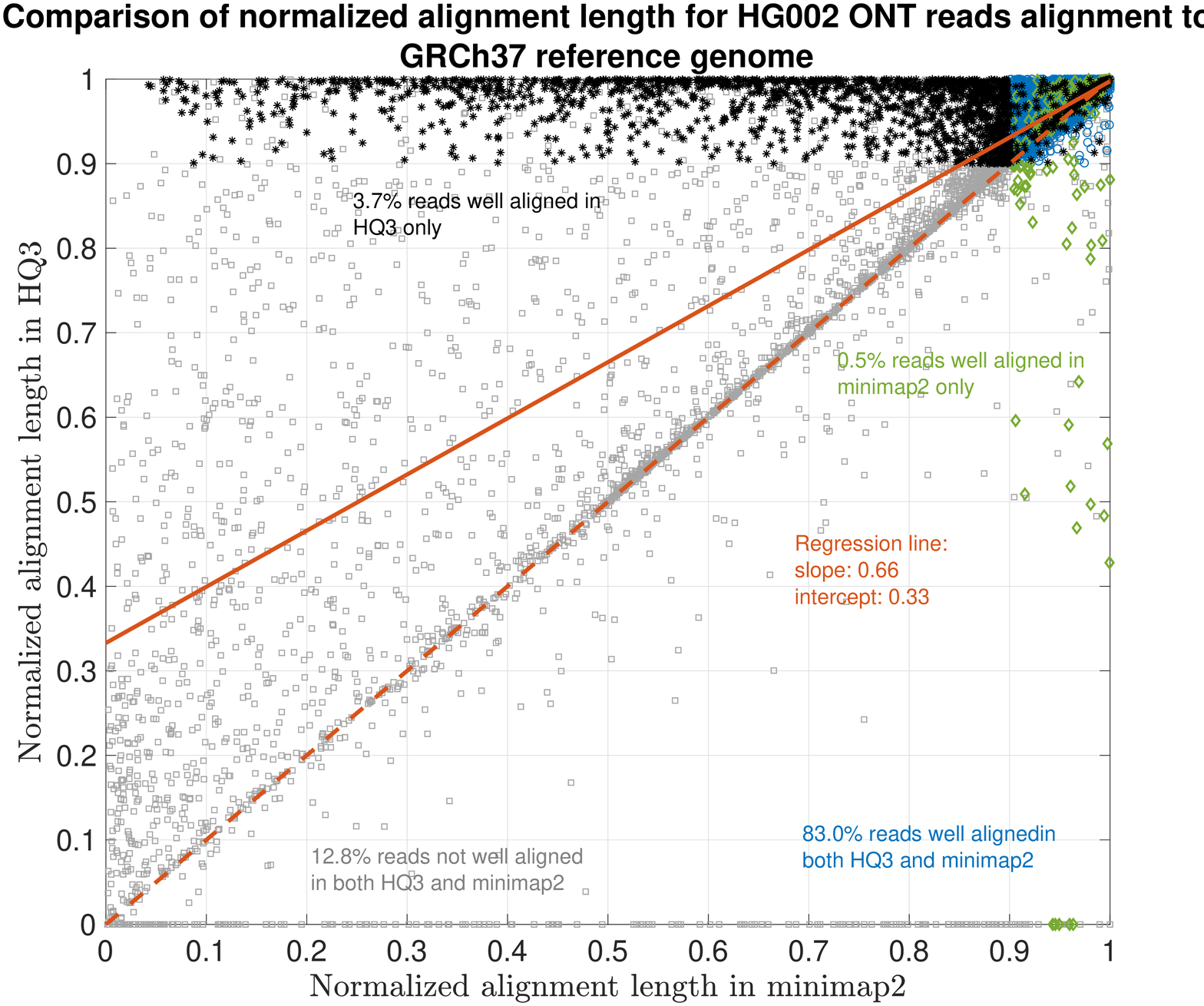}
		\label{fig:alignment_nal}
		(b)
	\end{minipage}\hfill
	\caption{{\bf HG002 nanopore long DNA reads alignment onto GRCh37 genome.}
	(a) Comparison of normalized edit distance for HG002 R9.4.1 PromethION reads data. Smaller values for normalized edit distance is desirable as it represents better alignment. The slope of the regression line is $0.82<1$, therefore, representing better alignments with $HQ3$ than minimap2 alignments for same reads on average.
	(b) Comparison of normalized alignment length for HG002 R9.4.1 PromethION reads data. Normalized alignment length of $1$ is desirable as it represents that entire read is aligned. The majority of the reads are above $y=x$ line representing longer alignment length in $HQ3$ than minimap2 alignment.}
	\label{fig:alignment}
\end{figure*}

\begin{figure*}[!h]
	\begin{minipage}{0.5\linewidth}
		\centering
		\includegraphics[width=\linewidth]{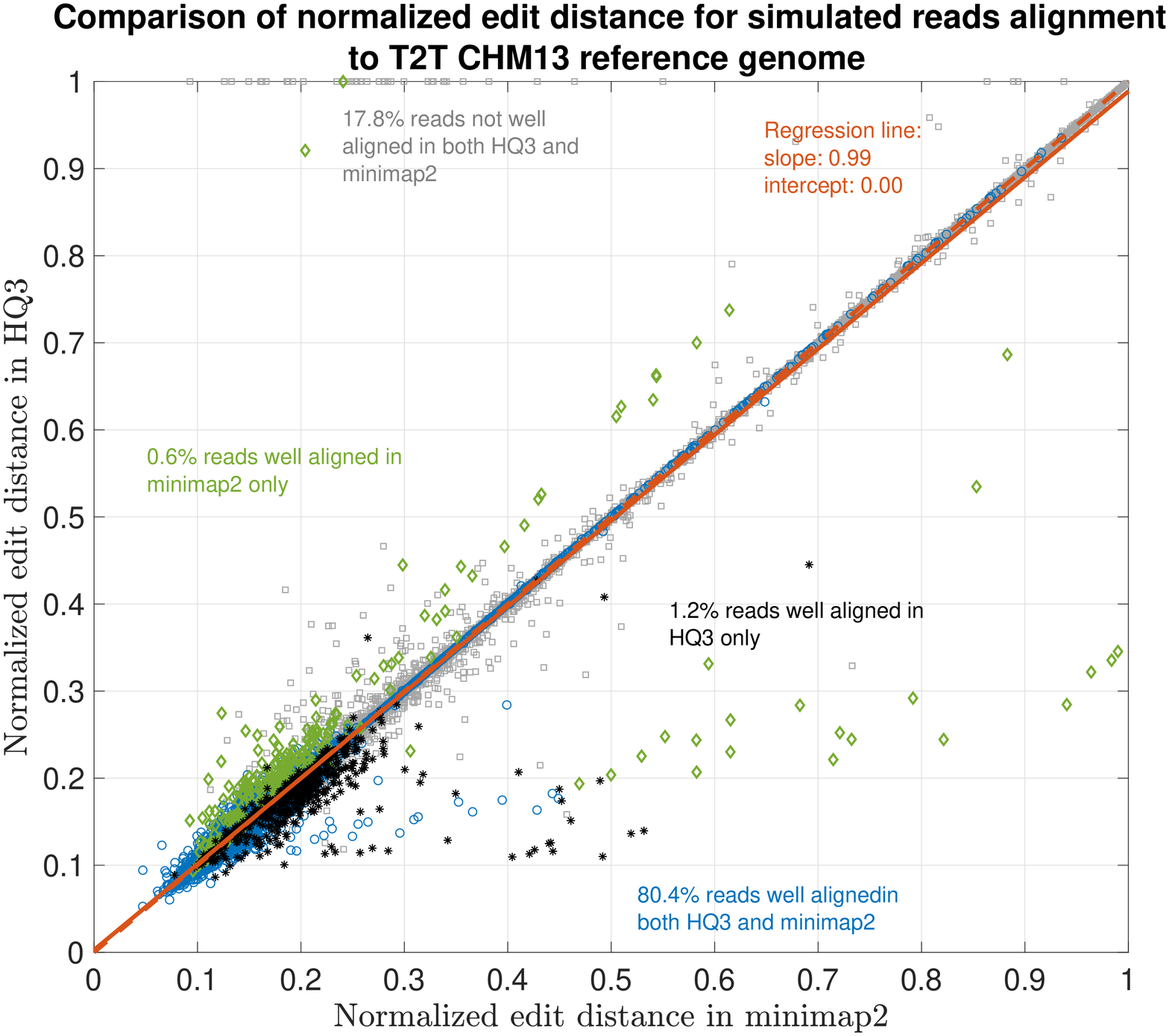}
		\label{fig:alignment_ned_sim_chm13}
		(a)
	\end{minipage}\hfill
	\begin{minipage}{0.5\linewidth}
		\centering
		\includegraphics[width=\linewidth]{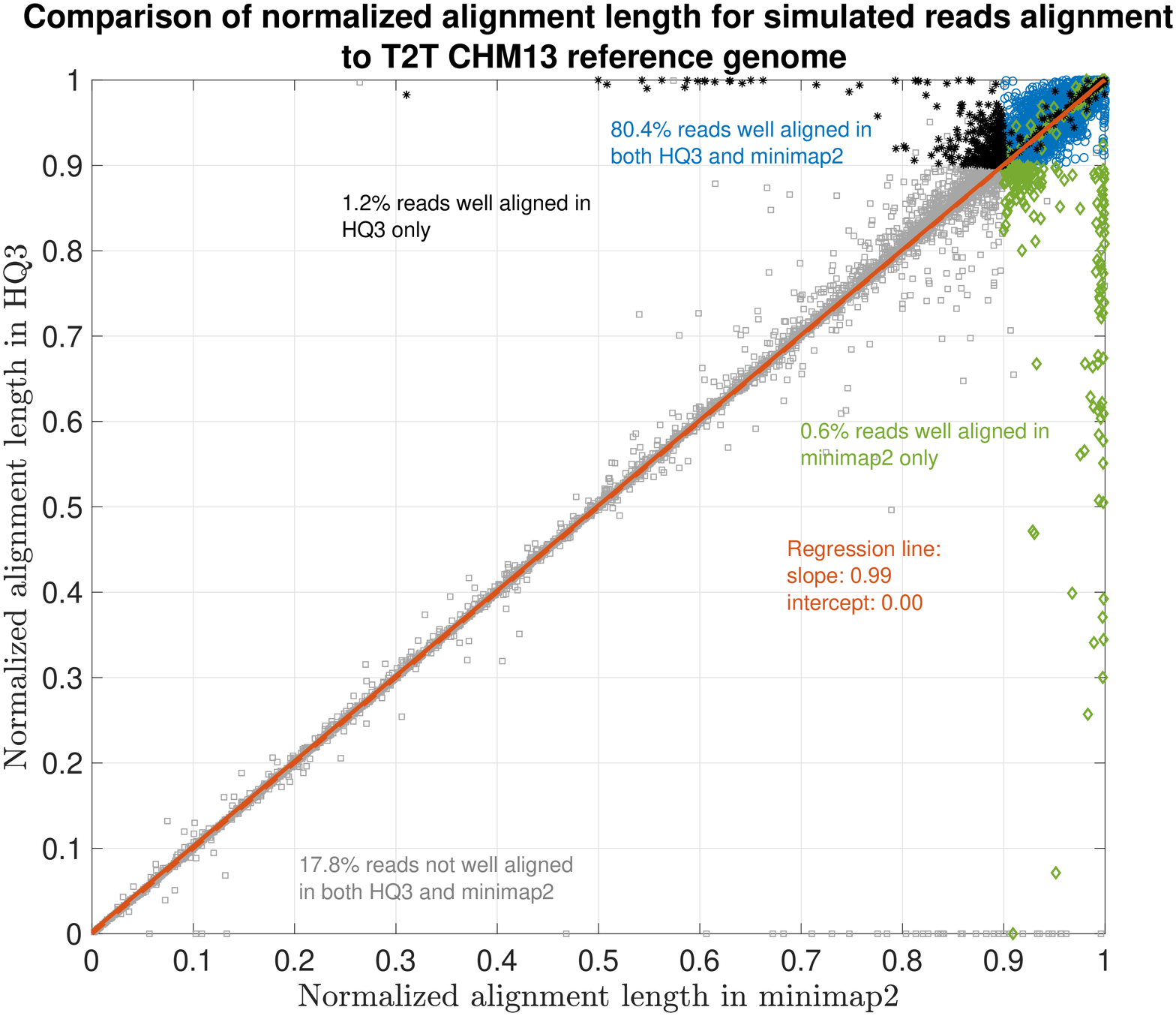}
		\label{fig:alignment_nal_sim_chm13}
		(b)
	\end{minipage}\hfill
	\caption{{\bf Simulated nanopore reads alignment onto T2T CHM13 genome.}
	(a) Comparison of normalized edit distance for simulated nanopore reads data. Smaller values for normalized edit distance is desirable as it represents better alignment. The slope of the regression line is $0.99<1$, therefore, representing marginally better alignments with $HQ3$ than minimap2 alignments for same reads on average.
	(b) Comparison of normalized alignment length for simulated nanopore reads data. Normalized alignment length of $1$ is desirable as it represents a contiguous alignment of the entire read.}
	\label{fig:alignment_sim_chm13}
\end{figure*}

\begin{table*}[!t]
\centering
\caption{Comparison for the percentage of well-aligned reads onto genome, and slope of the regression line (for normalized edit distance comparison plot of $HQ3$ vs minimap2 alignments) with randomly sampled reads for each datasets. The slope of the regression line shows the average gain in the normalized edit distance.}
\medskip
\begin{tabular}{|p{5.5cm}|p{2.3cm}|p{2.3cm}|p{2.3cm}|p{2.3cm}|}
\hline
\bf{Dataset (No. of sampled reads)} & \bf{Method of alignment} & \bf{Percentage well-aligned reads} & \bf{Slope of regression line} & \bf{Intercept}\\
\hline
\multirow{2}{=}{HG002 R9.4.1 reads to CHM13 (50k)}
  & minimap2
  & $85.64$
  & \multirow{2}{*}{$0.7940$}
  & \multirow{2}{*}{$0.0206$}
  \\ \cline{2-3}
  & $HQ3$
  & $89.35$
  & 
  & 
  \\ \cline{2-3}
  \hline
\multirow{2}{=}{HG002 R9.4.1 reads to GRCh37 (50k)} 
  & minimap2
  & $83.48$
  & \multirow{2}{*}{$0.8301$}
  & \multirow{2}{*}{$0.0181$}
  \\ \cline{2-3}
  & $HQ3$
  & $86.65$
  & 
  & 
  \\ \cline{2-3}
  \hline
\multirow{2}{=}{Simulated reads from chr 8 \& X of CHM13 assembly (50k)}
  & minimap2
  & $81.01$
  & \multirow{2}{*}{$0.9860$}
  & \multirow{2}{*}{$0.0028$}
  \\ \cline{2-3}
  & $HQ3$ 
  & $81.57$
  & 
  & 
  \\ \cline{2-3}
 \hline

\end{tabular}
\label{table:alignment}
\end{table*}


The alignment of DNA reads to the genome is a primitive step in structural variant calling pipelines \cite{DePristo2011}. $HQ3$ alignments shows an improvement over minimap2 alignments in terms of contiguity measured by normalized alignment length and alignment quality measured by normalized edit distance.

The results are illustrated in the Figures \ref{fig:alignment_chm13}, \ref{fig:alignment}, \ref{fig:alignment_sim_chm13}, and Table \ref{table:alignment}. At a coarse level, the performance is measured by the fraction of the reads that are well-aligned by the algorithm. A read is  well-aligned if at-least $90\%$ of the read is aligned to genome and has a high mapping quality (see Methods). HQAlign improves the fraction of well-aligned reads than minimap2 - in particular, in the HG002 R9.4.1 reads alignment to T2T CHM13 reference, this metric improves to $89.35\%$ from $85.64\%$, and for the alignments to GRCh37 reference, this metric improves to $86.65\%$ from $83.48\%$. Furthermore, there are $310,036$ reads with at-least 1kb additional bases aligned using HQAlign compared to minimap2 alignments for T2T CHM13 reference, and there are $299,896$ reads with at-least 1kb additional bases aligned using HQAlign compared to minimap2 for GRCh37 reference.

The results in Figure \ref{fig:alignment_chm13} and \ref{fig:alignment} compares the quality of the alignments using minimap2 and HQAlign at a fine-grained level for HG002 ONT reads alignment to T2T CHM13 genome and GRCh37 genome, respectively. Figure \ref{fig:alignment_chm13}a and \ref{fig:alignment}a compares the normalized edit distance for HQAlign and minimap2.
The normalized edit distance is the edit distance between the entire read and the aligned section on the genome normalized by the length of the read, in nucleotide domain for \emph{both} minimap2 alignment and quantized alignment ($HQ3$). In case of $HQ3$, the information of the location of the alignment on the genome is leveraged from the quantized read and the quantized genome alignment, and the edit distance is computed between the corresponding nucleotide read and the aligned region on the nucleotide genome (see Methods for details). Intuitively, normalized edit distance gives a measure of how close the two sequences are. Therefore, the smaller the normalized edit distance, better is the alignment.


Figure \ref{fig:alignment_chm13}a shows that for alignments of the reads to T2T CHM13 reference, the normalized edit distance is on average smaller for $HQ3$ alignments than minimap2 alignments. The better alignment in $HQ3$ is also evident from the slope of the regression line in Figure \ref{fig:alignment_chm13}a. It shows that on average $HQ3$ alignments has $21\%$ improvement in terms of the normalized edit distance than the minimap2 alignments. Well aligned reads in both $HQ3$ and minimap2 are represented by blue circles in Figure \ref{fig:alignment_chm13}, well aligned reads in $HQ3$ only are represented in black asterisks, well aligned in minimap2 only are represented in green diamonds and reads that are not well aligned in both are represented in grey squares.
Further, it is important to note that for normalized edit distance less than $0.1$, the alignments are marginally better in the DNA space, but for normalized edit distance higher than $0.1$, the alignments are significantly better in $HQ3$ space, especially, the $4\%$ reads that are well aligned in $HQ3$ and not well aligned in minimap2. This is because of higher contiguity of alignments in $HQ3$ space and signifies the improvement by $HQ3$ when the error rates is higher. For alignments to GRCh37 reference, $HQ3$ has an average improvement of $17\%$, as shown in Figure \ref{fig:alignment}a.

The results for another fine-grained metric are shown in Figure \ref{fig:alignment_chm13}b and \ref{fig:alignment}b, which compares the normalized alignment length in $HQ3$ to the normalized alignment length in minimap2 alignments. The normalized alignment length is the ratio of the length of the section on genome where a read aligns to the length of the read. In Figure \ref{fig:alignment_chm13}b, there are $4\%$ reads that are well-aligned in $HQ3$ only, and the normalized alignment length is close to 1 in $HQ3$ but it is much less than 1 in minimap2, therefore representing several non-contiguous alignments in nucleotide domain that are captured as contiguous alignment in $HQ3$. In Figure \ref{fig:alignment}b, there are $3.7\%$ that are well-aligned in $HQ3$ only.

We have also benchmarked the performance of HQAlign with the simulated reads data and compared its alignment performance with minimap2 in Figure \ref{fig:alignment_sim_chm13}. The ONT reads are simulated from chromosome 8 and X of CHM13 T2T assembly using nanosim \cite{yang2017nanosim} with a coverage of 40x, median and mean read length $4.5$ kb and $14$ kb, respectively. The results shows that the alignment performance of both HQAlign and minimap2 are at par with each other.

\subsection{SV calling}

\subsubsection{Dataset}
Long read sequencing plays an important role in detecting structural variations. We evaluated SV detection using minimap2 and HQAlign with Sniffles2 as the variant calling algorithm on both real and simulated data. We simulated 2000 INDELS and 200 Inversion SVs on chromosome 8 and X of T2T CHM13 reference genome using SURVIVOR \cite{jeffares2017transient} with SV length uniformly distributed between 50 and 10000, and the ONT reads are simulated using nanosim with average length of 14k, median length of 4.5k and maximum length 2.5Mbp at a coverage of 40x. We have used Truvari to benchmark the calls against the truth set. For real data alignment with GRCh37 as the reference genome, the SV calls are compared against the ground truth sets from (1) Genome In A Bottle (GIAB) Tier 1 calls \cite{zook2020robust} and (2) another truth set is constructed by comparing the haplotype-resolved assembly of HG002 against GRCh37 reference genome using dipcall \cite{li2018synthetic}. For T2T CHM13 reference genome, since the ground truth for SVs is not available, we have constructed the truth set by comparing the haplotype-resolved assembly of HG002 against CHM13 reference using dipcall.
However, it is hard to establish ground truth for the SV calls that are made in the centromere regions, even though the assembly is likely to be correct. Therefore, we have provided both the analysis including the SV calls in centromere regions (in Figures \ref{fig:complimentary_chm13} and \ref{fig:sv_chm13}) and the analysis for SV calls excluding the centromere regions (in Figures \ref{fig:complimentary_chm13-cen} and \ref{fig:sv_chm13-centromere}).

\subsubsection{SV calling results}

\begin{figure}[!h]
	\begin{minipage}{0.5\linewidth}
		\centering
		\includegraphics[width=0.7\linewidth]{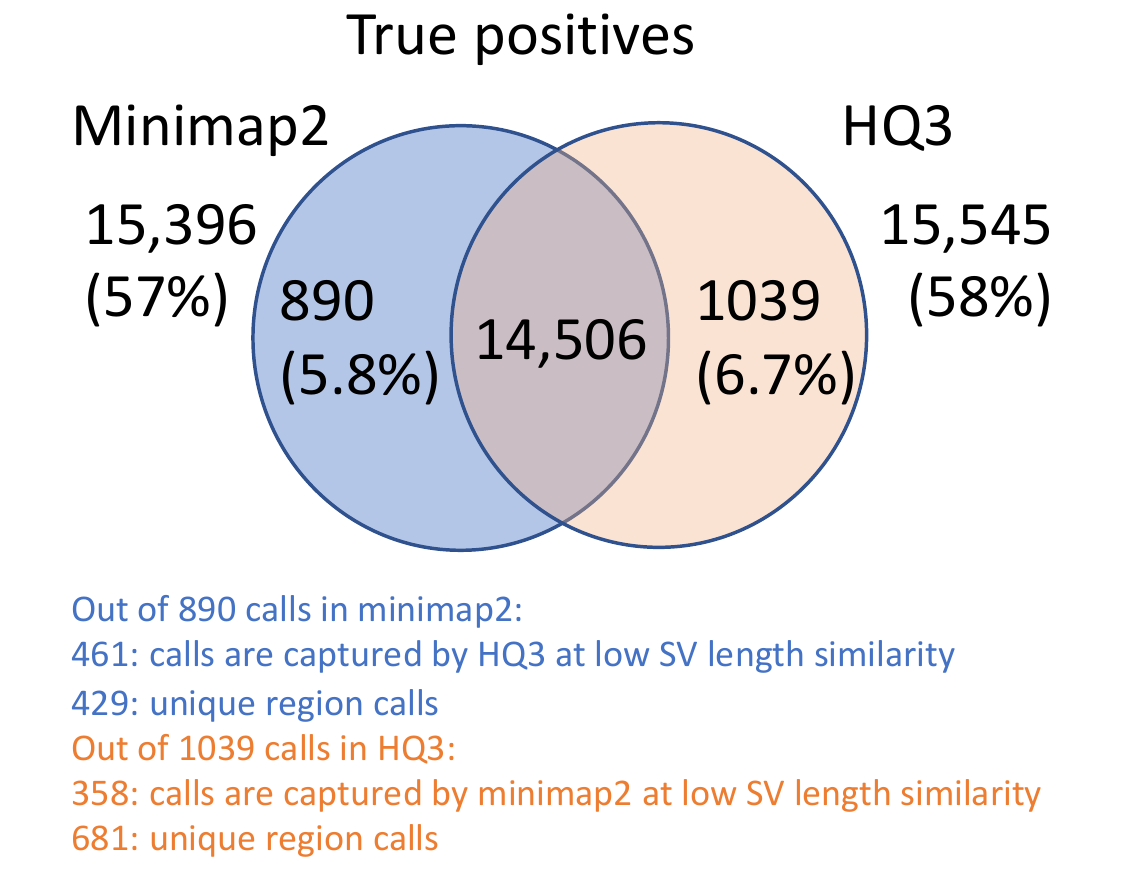}\\
		\label{fig:tp_chm13}
		\centering
		(a)
	\end{minipage}
	\begin{minipage}{0.5\linewidth}
		\centering
		\includegraphics[width=0.7\linewidth]{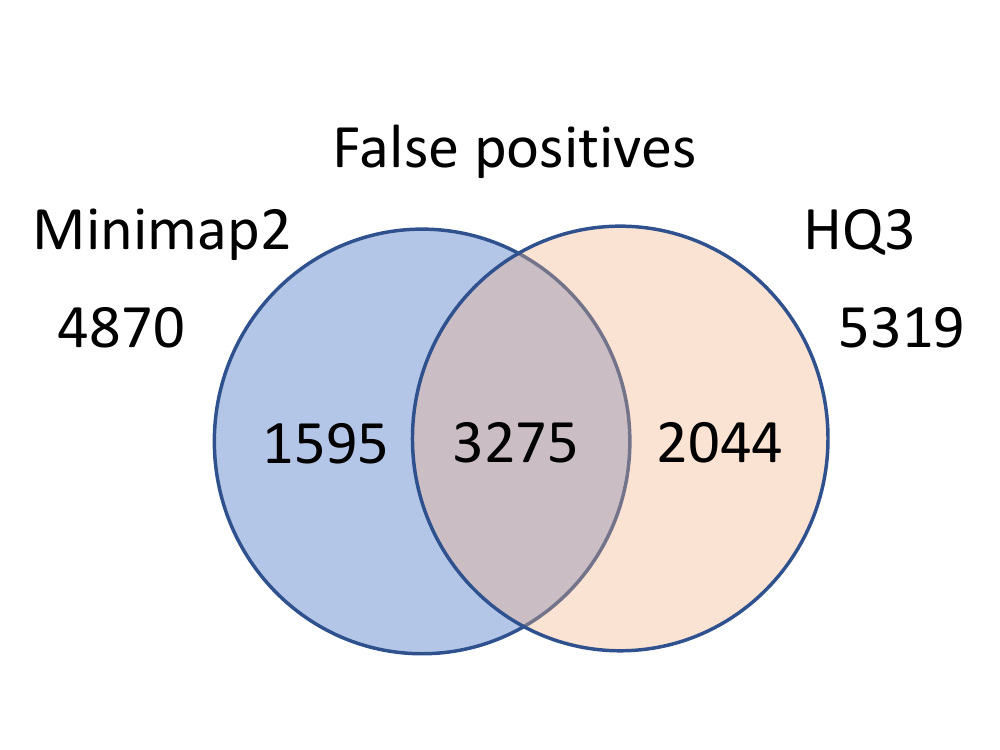}\\
		\label{fig:fp_chm13}
		\centering
		(b)
	\end{minipage}\hfill
	\caption{{\bf Comparison of SV calls from $HQ3$ and minimap2 with HG002-to-CHM13 dipcall as truth set.}
	(a) Comparison of true positive calls.
	(b) Comparison of false positive calls.}
	\label{fig:complimentary_chm13}
\end{figure}

\begin{figure*}[!h]
	\begin{minipage}{0.5\linewidth}
		\centering
		\includegraphics[width=0.7\linewidth]{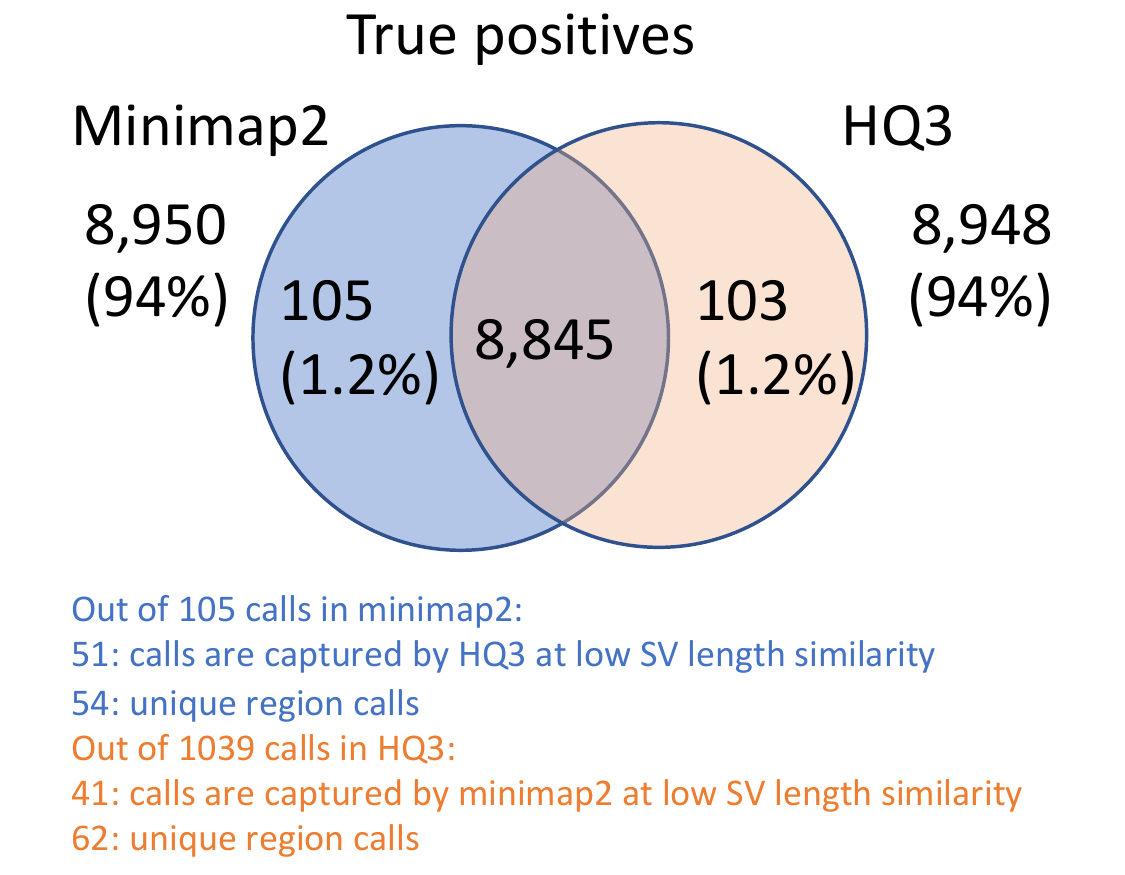}\\
		\label{fig:tp_giab}
		\centering
		(a)
	\end{minipage}\hfill
	\begin{minipage}{0.5\linewidth}
		\centering
		\includegraphics[width=0.7\linewidth]{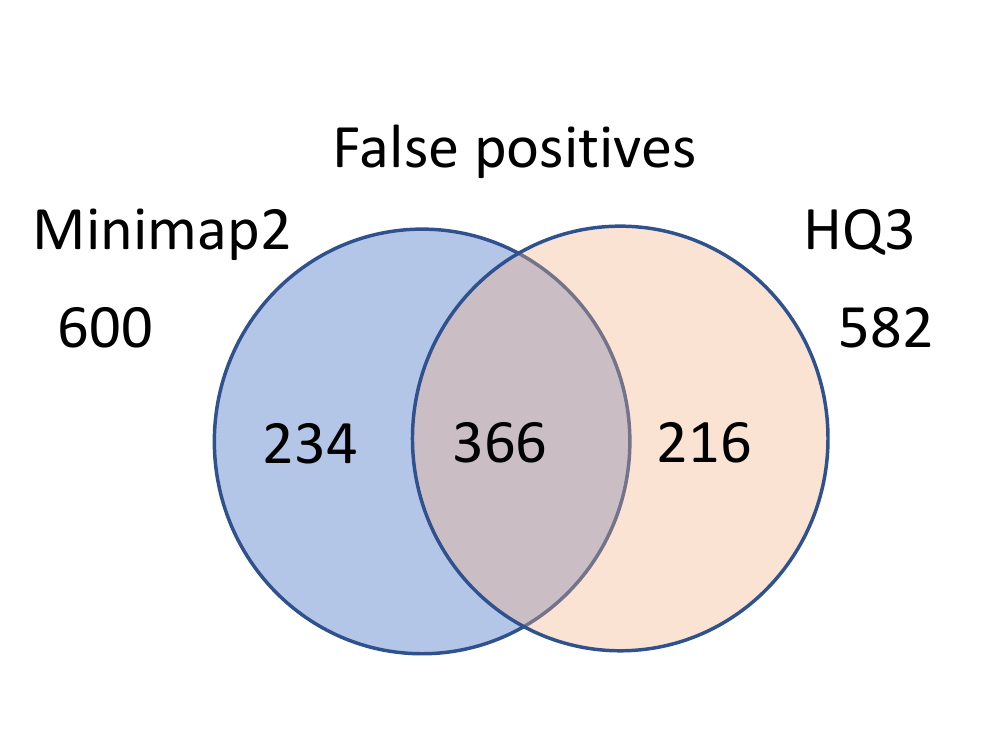}\\
		\label{fig:fp_giab}
		\centering
		(b)
	\end{minipage}\hfill
	\caption{{\bf Comparison of SV calls from $HQ3$ and minimap2 with Genome in a Bottle (GIAB) Tier 1 truth set for GRCh37 build.}
	(a) Comparison of true positive calls.
	(b) Comparison of false positive calls.}
	\label{fig:complimentary_giab}
\end{figure*}

\begin{figure*}[!h]
	\begin{minipage}{0.5\linewidth}
		\centering
		\includegraphics[width=0.7\linewidth]{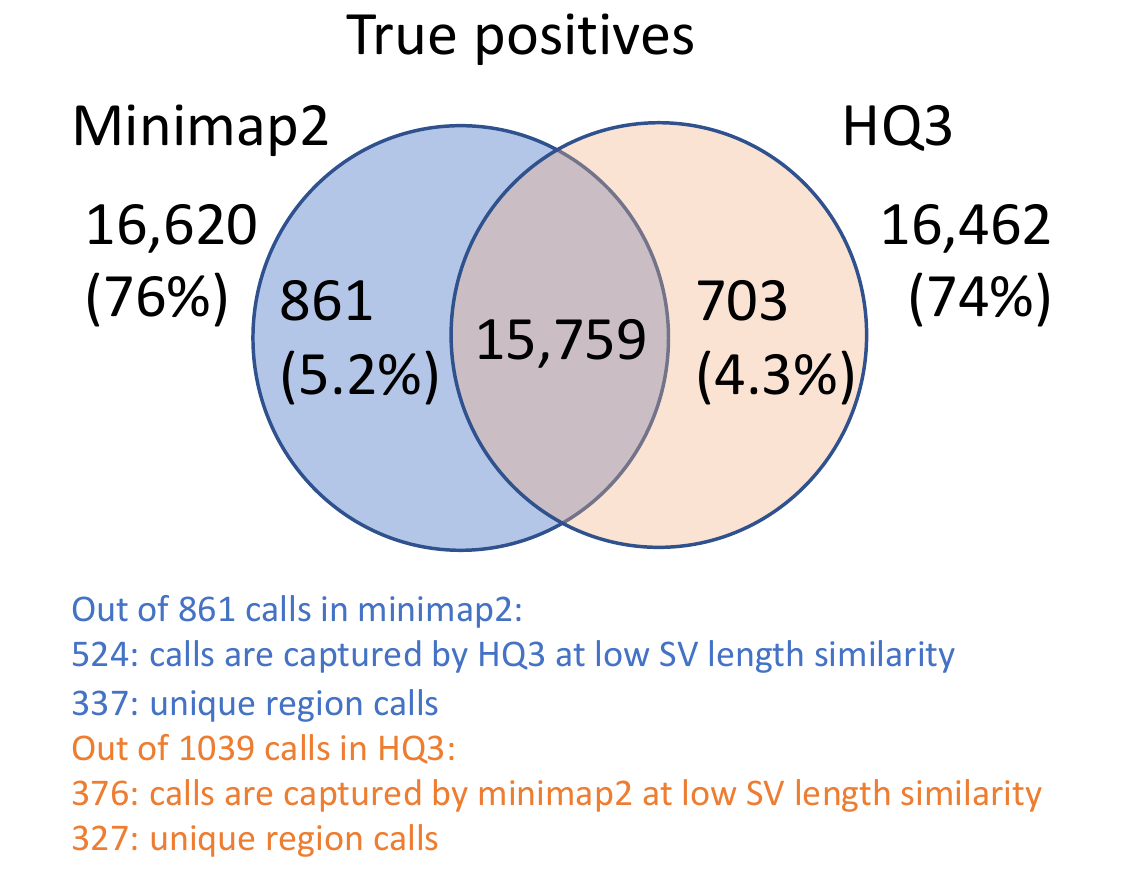}\\
		\label{fig:tp_hg002}
		(a)
	\end{minipage}\hfill
	\begin{minipage}{0.5\linewidth}
		\centering
		\includegraphics[width=0.7\linewidth]{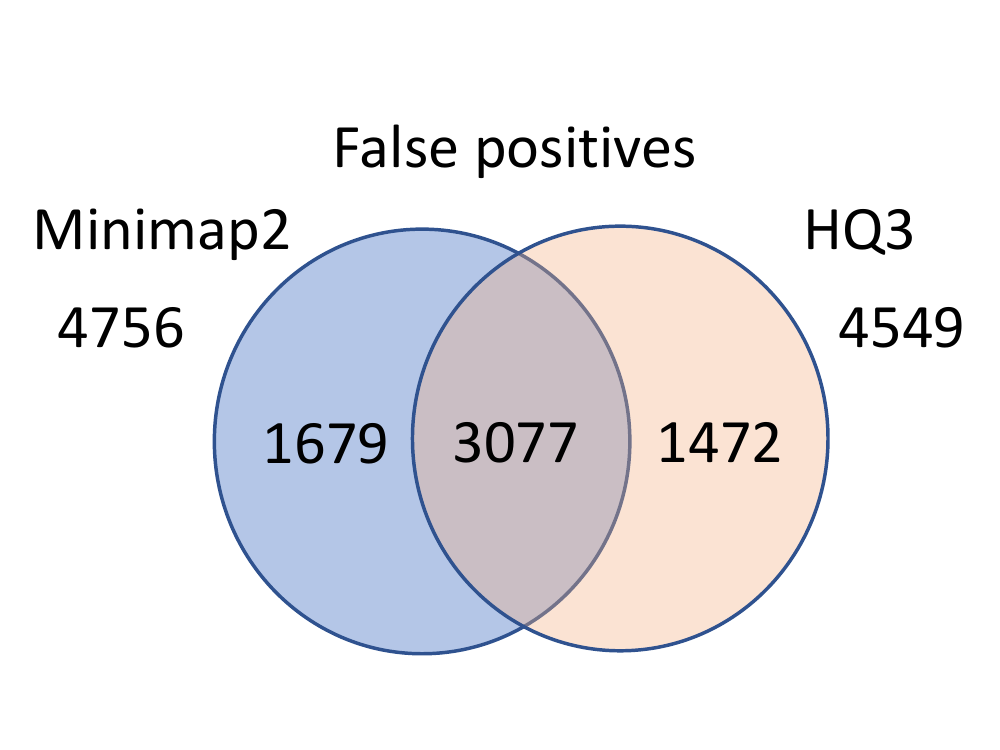}\\
		\label{fig:fp_hg002}
		(b)
	\end{minipage}\hfill
	\caption{{\bf Comparison of SV calls from $HQ3$ and minimap2 with HG002-to-GRCh37 dipcall as truth set.}
	(a) Comparison of true positive calls.
	(b) Comparison of false positive calls.}
	\label{fig:complimentary_hg002}
\end{figure*}

\begin{figure*}[!h]
	\begin{minipage}{0.5\linewidth}
		\centering
		\includegraphics[width=0.7\linewidth]{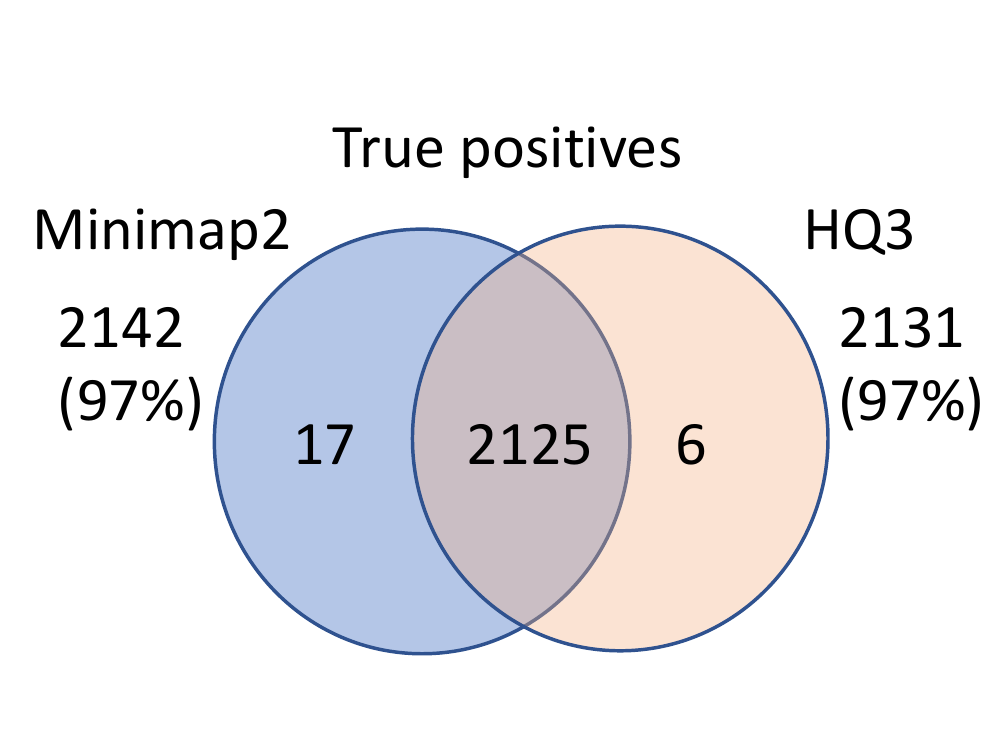}\\
		\label{fig:tp_sim}
		(a)
	\end{minipage}\hfill
	\begin{minipage}{0.5\linewidth}
		\centering
		\includegraphics[width=0.7\linewidth]{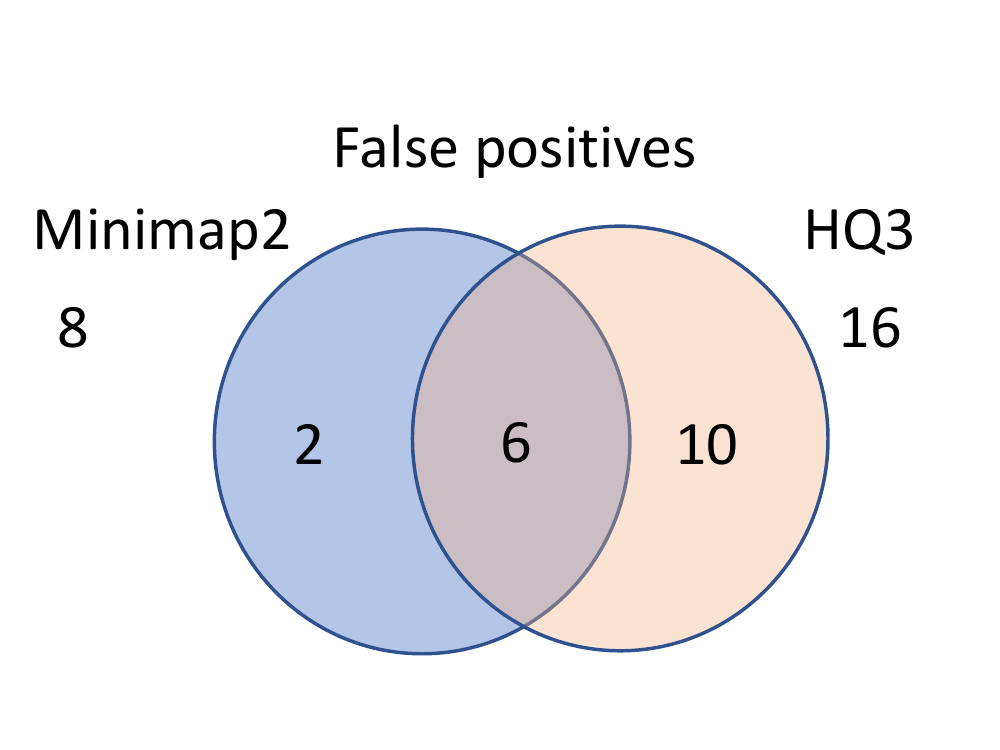}\\
		\label{fig:fp_sim}
		(b)
	\end{minipage}\hfill
	\caption{{\bf Comparison of SV calls from $HQ3$ and minimap2 with simulated data.}
	(a) Comparison of true positive calls.
	(b) Comparison of false positive calls.}
	\label{fig:complimentary_sim}
\end{figure*}

\begin{figure*}[!h]
	\begin{minipage}{0.5\linewidth}
		\centering
		\includegraphics[width=0.7\linewidth]{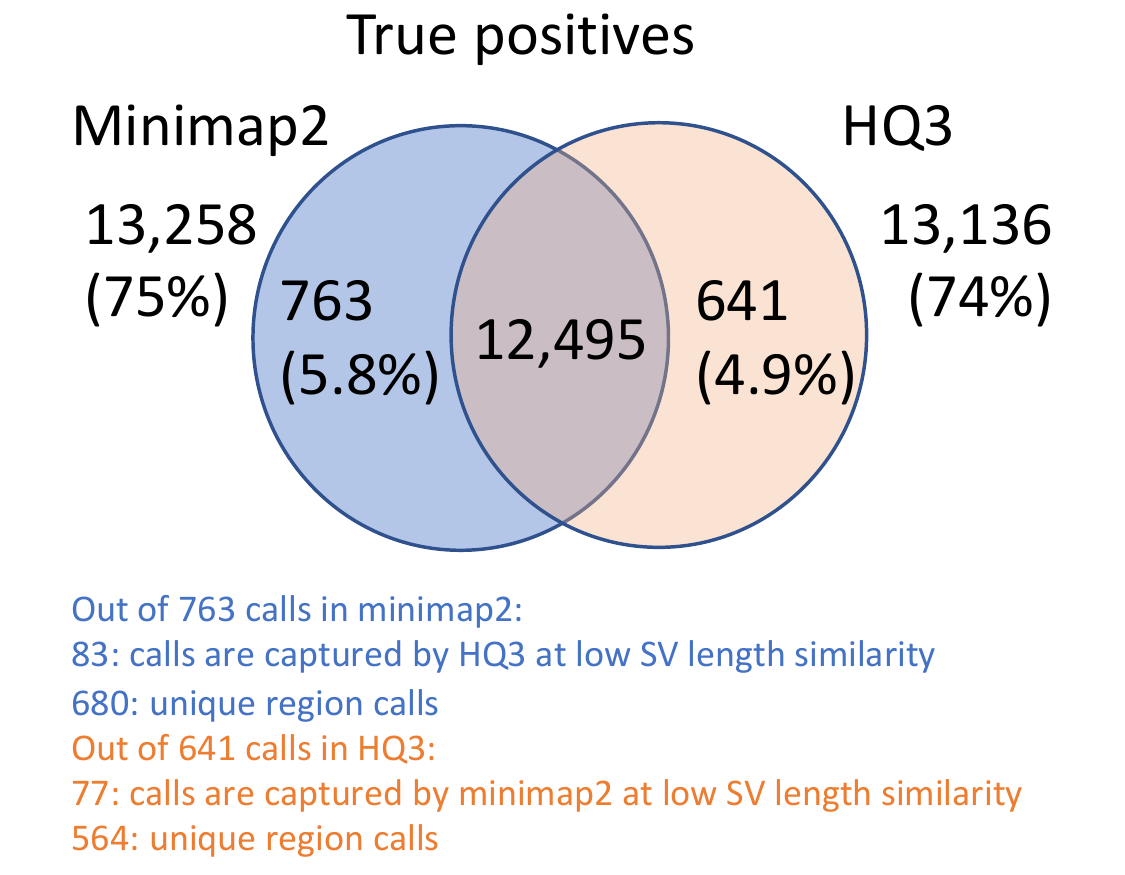}\\
		\label{fig:tp_chm13-cen}
		(a)
	\end{minipage}\hfill
	\begin{minipage}{0.5\linewidth}
		\centering
		\includegraphics[width=0.7\linewidth]{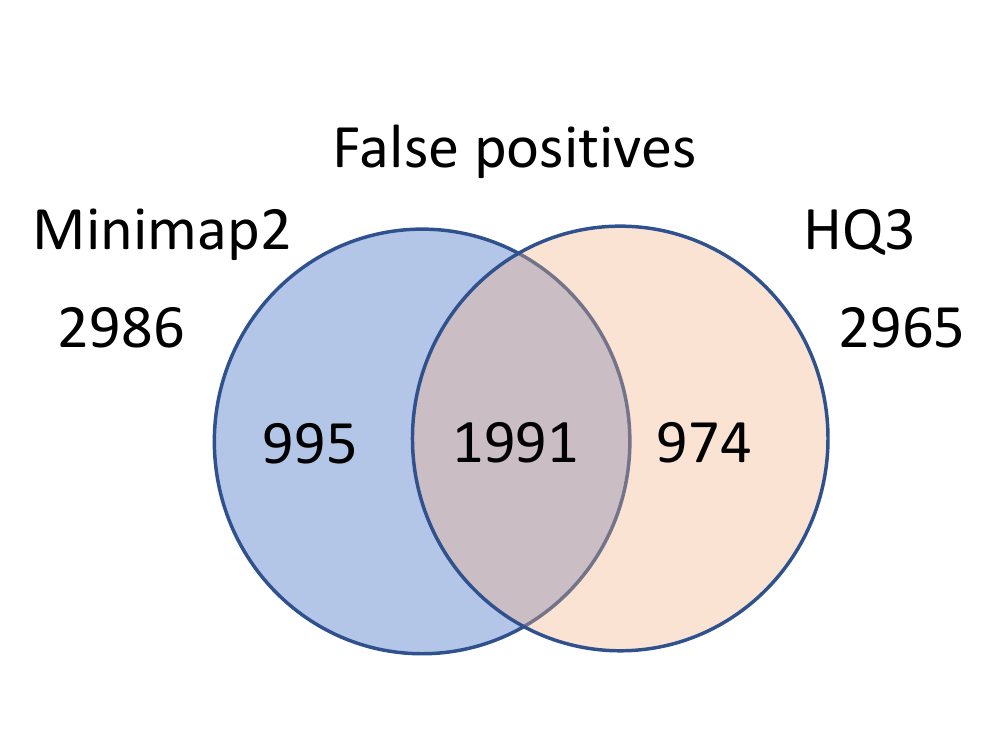}\\
		\label{fig:fp_chm13-cen}
		(b)
	\end{minipage}\hfill
	\caption{{\bf Comparison of SV calls from $HQ3$ and minimap2 with HG002-to-CHM13 dipcall as truth set and excluding the calls from the centromere regions.}
	(a) Comparison of true positive calls.
	(b) Comparison of false positive calls.}
	\label{fig:complimentary_chm13-cen}
\end{figure*}

The standalone performance of both $HQ3$ and minimap2 is at par with each other across different references and truth set used in this study for real data as well as for the simulated data in terms of the F1 score. However, both $HQ3$ and minimap2 detects complementary SV calls most likely in the repeat regions where accurate alignment is difficult and therefore, leads to many broken calls.

The analysis with comparison of SV calls from $HQ3$ and minimap2 with GIAB Tier 1 truth set gives a precision, recall and F1 score of $0.94$, $0.94$, and $0.94$, respectively for both minimap2 and $HQ3$. A union model of minimap2 and $HQ3$ can improve the recall rate at the same F1 score, and the union model has a precision, recall, and F1 score of $0.93$, $0.95$, and $0.94$, respectively. Moreover, out of $103$ SV calls that are made by $HQ3$ only (Figure \ref{fig:complimentary_giab}), $41$ calls are made by minimap2 alignments at a lower SV length similarity and $62$ calls are unique region calls. Out of $105$ SV calls made by minimap2 only, $51$ are captured by $HQ3$ at a lower SV length similarity and $54$ are unique region calls. $HQ3$ improves the breakpoint accuracy for $14.11\%$ calls that have higher than $50$ difference in breakpoints and it improves the length similarity of $19.97\%$ calls that have SV length similarity lower than $0.95$ (Figure \ref{fig:sv_giab}).

\begin{figure*}[!h]
	\begin{minipage}{0.5\linewidth}
		\centering
		\includegraphics[width=\linewidth]{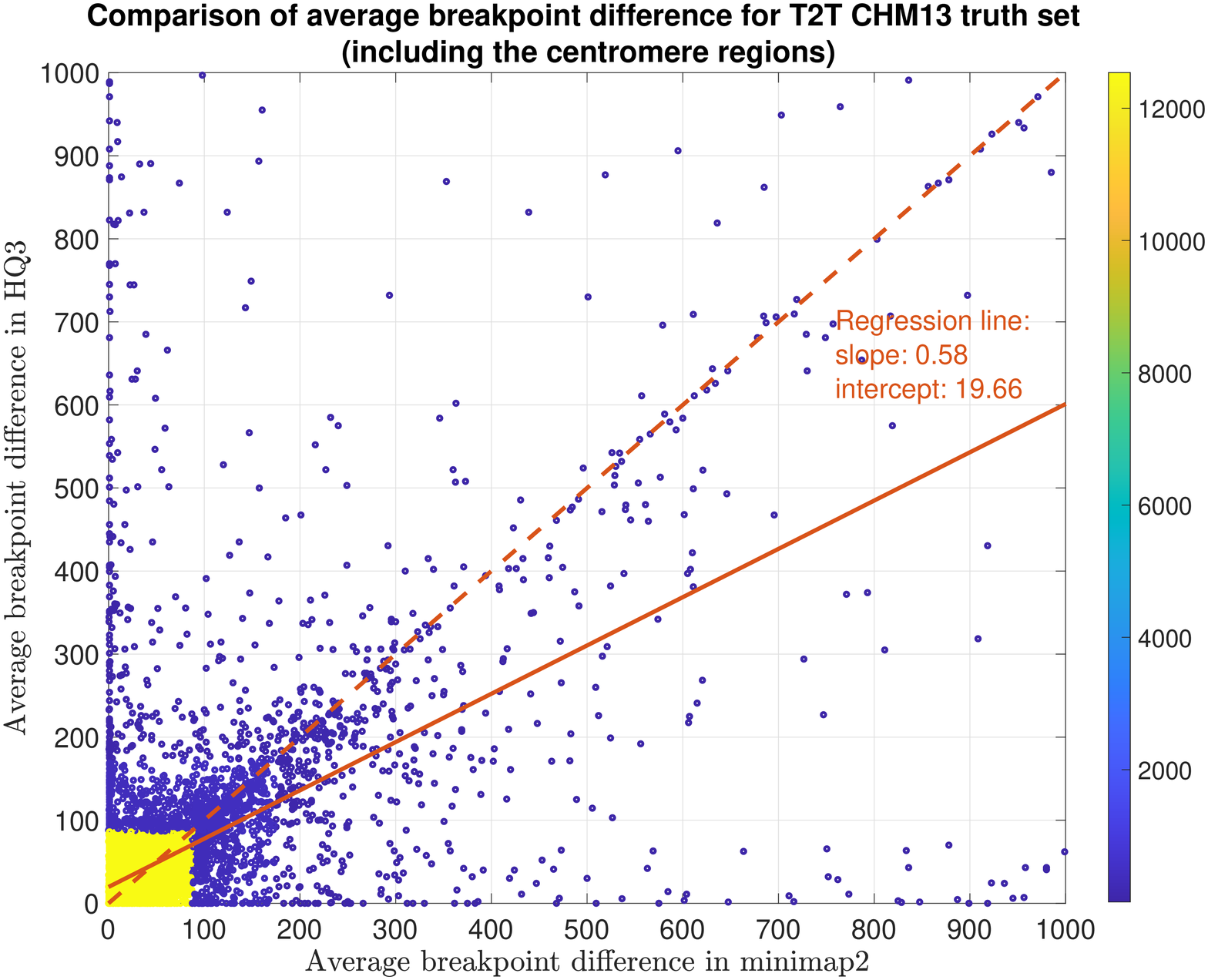}
		\label{fig:sv_chm13_breakpoints}
		(a)
	\end{minipage}\hfill
	\begin{minipage}{0.5\linewidth}
		\centering
		\includegraphics[width=\linewidth]{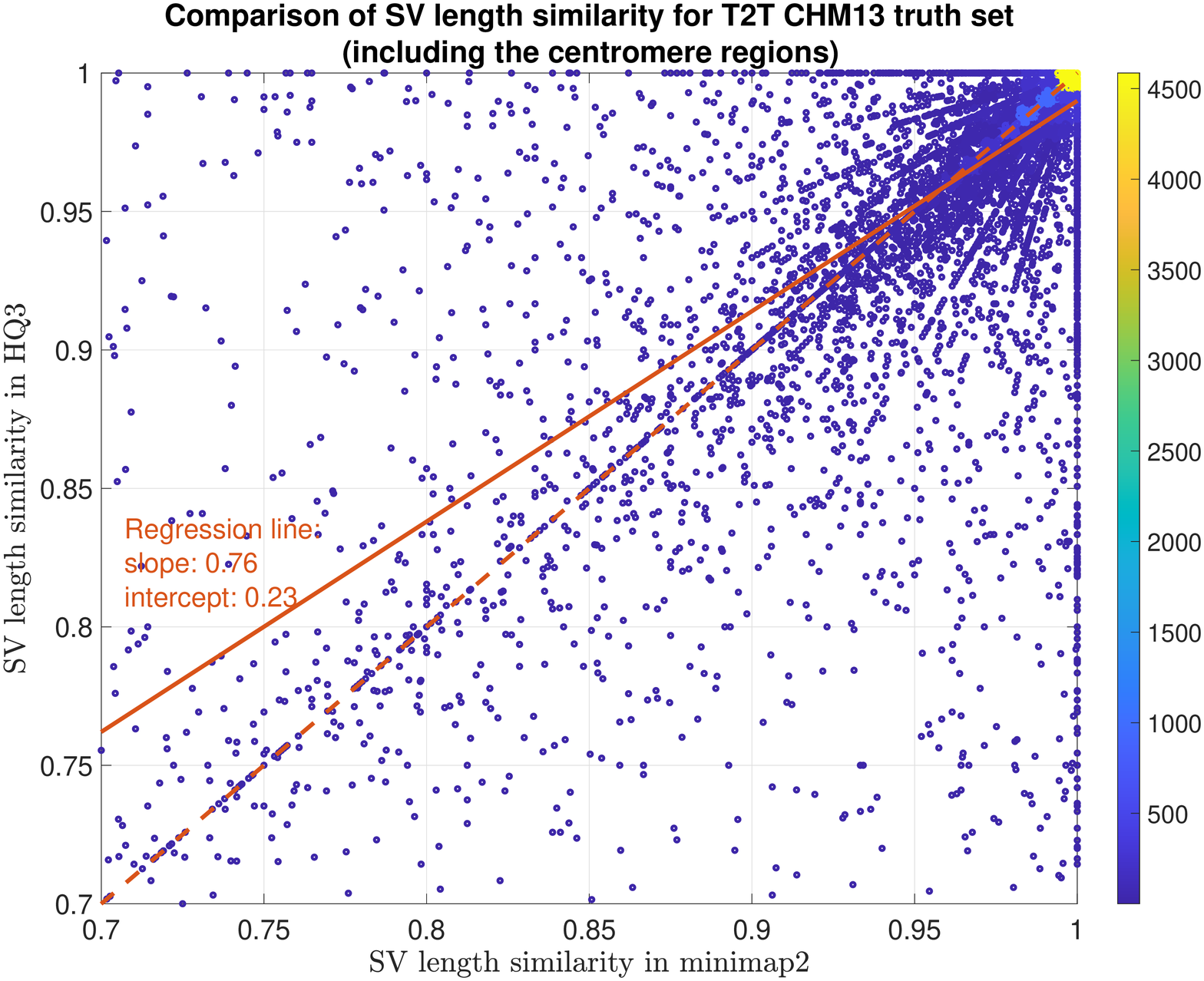}
		\label{fig:sv_chm13_svlength}
		(b)
	\end{minipage}\hfill
	\caption{{\bf SV quality comparison for common true positive calls in $HQ3$ and minimap2 against HG002-to-CHM13 dipcall truth set.}
	(a) Comparison of SV breakpoint accuracy in $HQ3$ and minimap2 for common true positive calls. The difference of SV breakpoint is compared to the truth set generated from comparing HG002 haplotype-resolved assembly to T2T CHM13 build. A smaller difference represents better breakpoint accuracy. Therefore, slope of the regression line $0.58<1$ represents better accuracy of $HQ3$ than minimap2 on average.
	(b) Comparison of SV length similarity in $HQ3$ and minimap2 for common true positive calls. The slope of the regression line $0.76<1$ represents better SV length in minimap2 than $HQ3$ on average, but the intercept is high ($0.23$). However, this is due to a large density of SVs with length similarity $\geq 0.95$ in both minimap2 and $HQ3$. For length similarity less than $0.95$, $HQ3$ has better performance than minimap2.}
	\label{fig:sv_chm13}
\end{figure*}

\begin{figure*}[!h]
	\begin{minipage}{0.5\linewidth}
		\centering
		\includegraphics[width=\linewidth]{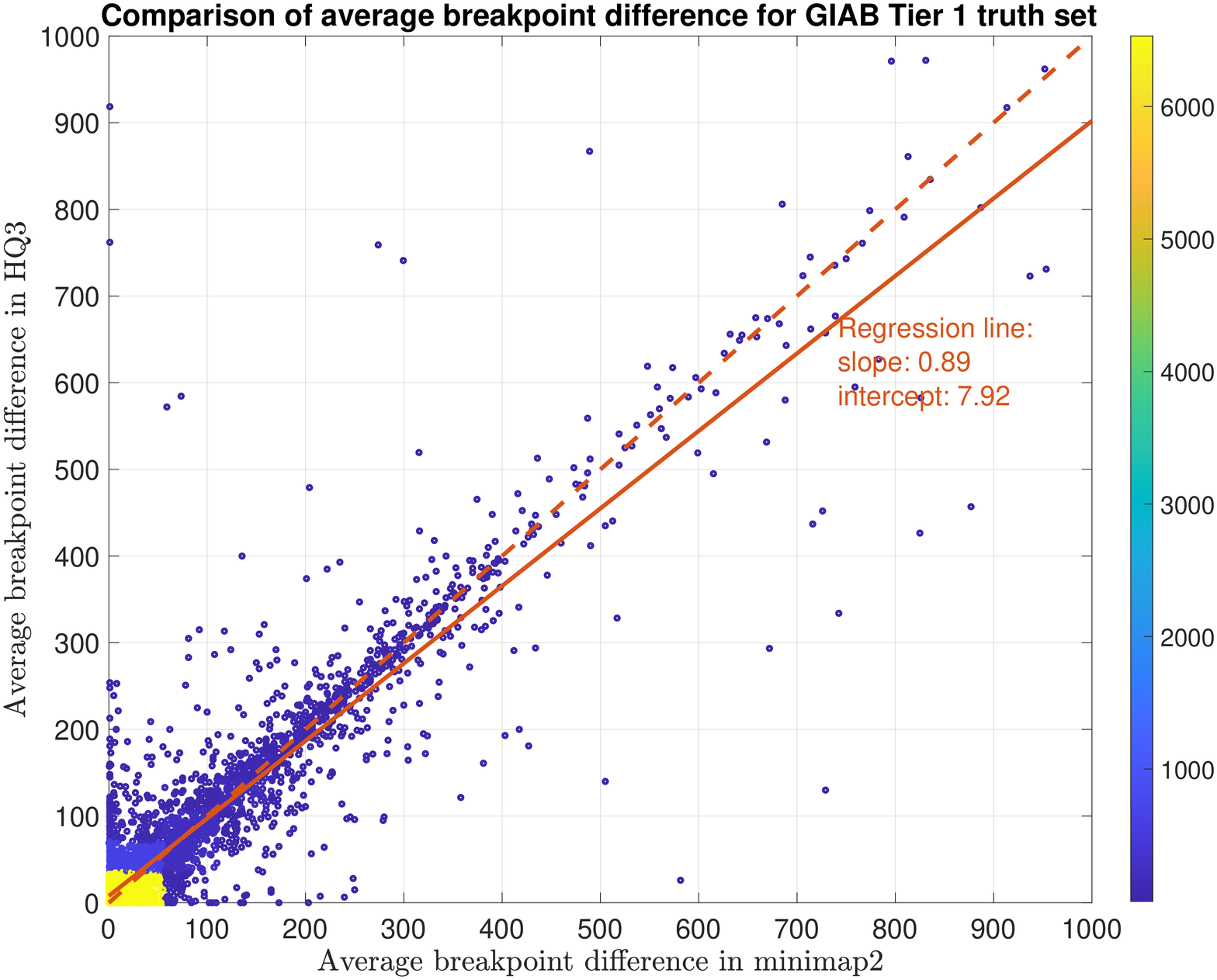}
		\label{fig:sv_giab_breakpoints}
		(a)
	\end{minipage}\hfill
	\begin{minipage}{0.5\linewidth}
		\centering
		\includegraphics[width=\linewidth]{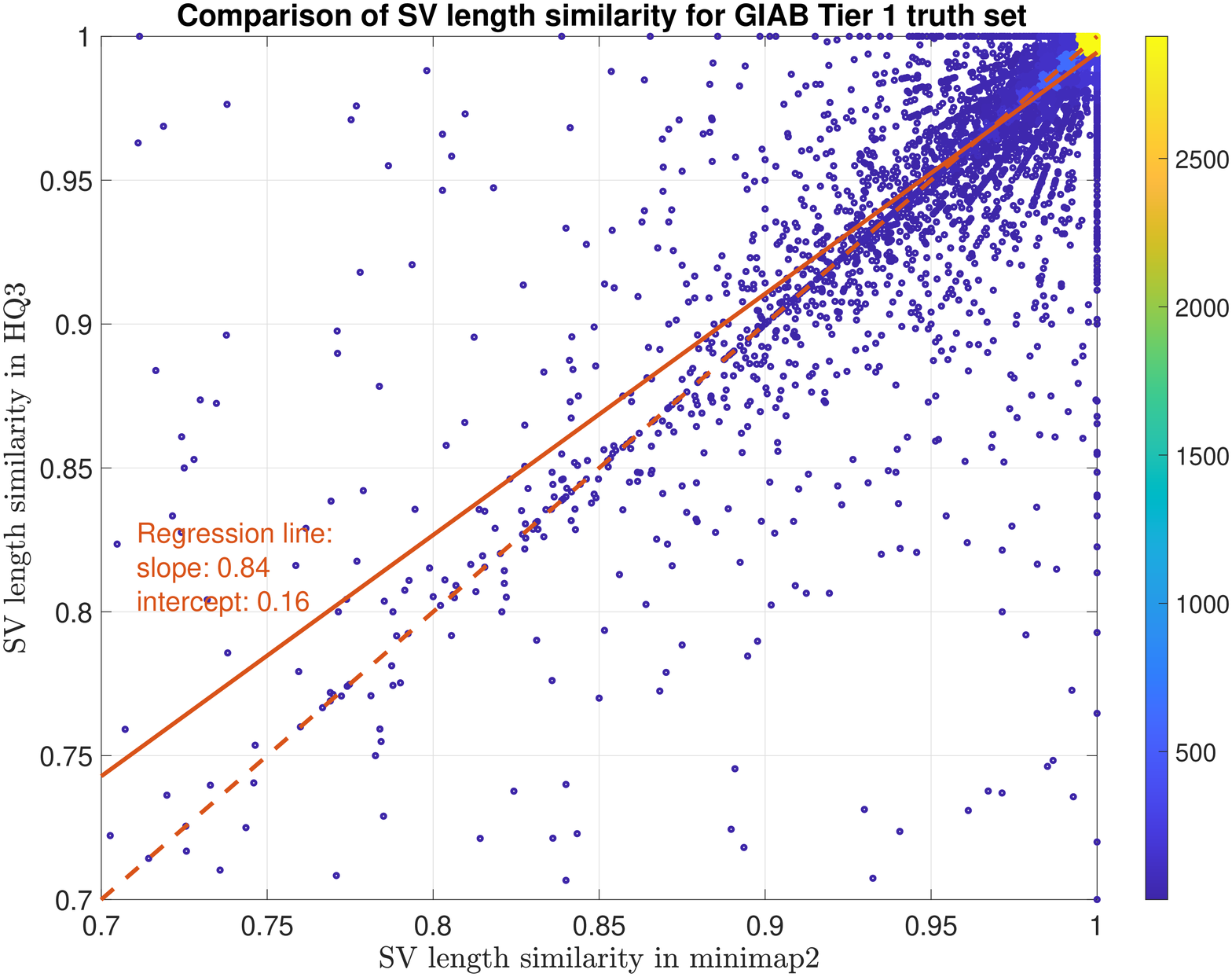}
		\label{fig:sv_giab_svlength}
		(b)
	\end{minipage}\hfill
	\caption{{\bf SV quality comparison for common true positive calls in $HQ3$ and minimap2 against GIAB Tier 1 truth set.}
	(a) Comparison of SV breakpoint accuracy in $HQ3$ and minimap2 for common true positive calls. The difference of SV breakpoint is compared to the GIAB Tier 1 truth set. A smaller difference represents better breakpoint accuracy. Therefore, slope of the regression line $<1$ represents better accuracy of $HQ3$ than minimap2 on average.
	(b) Comparison of SV length similarity in $HQ3$ and minimap2 for common true positive calls. The slope of the regression line $<1$ represents better SV length in minimap2 than $HQ3$ on average. However, this is due to a large density of SVs with length similarity $\geq 0.95$ in both minimap2 and $HQ3$. For length similarity less than $0.95$, $HQ3$ has better performance than minimap2.}
	\label{fig:sv_giab}
\end{figure*}

We have compared the SV calls made by HG002 reads against T2T CHM13 reference genome using both minimap2 and $HQ3$ and benchmarking them against truth set generated by comparing HG002 haplotype-resolved assembly to T2T CHM13 assembly. The standalone performance have precision, recall and F1 score of $0.77$, $0.57$ and $0.66$, respectively for minimap2 and $0.75$, $0.58$ and $0.65$, respectively for $HQ3$. However, because of high number of complementary true positive calls in minimap2 and $HQ3$, the union model has a significant improved recall at the same F1 score with precision, recall and F1 score of $0.71$, $0.61$ and $0.66$, respectively. Out of $1039$ ($6.7\%$) calls that are made in $HQ3$ only, $358$ are captured by minimap2 at a lower SV length similarity threshold and $681$ are unique calls, whereas out of $890$ ($5.8\%$) calls that are made by minimap2 only, $461$ are captured by $HQ3$ at a lower SV length similarity threshold and $429$ are unique (as shown in Figure \ref{fig:complimentary_chm13}a).
Further, for the common true positive calls in both minimap2 and $HQ3$, we observe a similar pattern as the other datasets in improvement of breakpoint accuracy with $HQ3$ for $18.66\%$ calls that have difference in breakpoint greater than $50$, and improvement in SV length similarity for $19.76\%$ calls with similarity less than $0.95$ (Figure \ref{fig:sv_chm13}a-b).

\begin{table*}[!h]
\centering
\caption{Comparison for precision, recall and and F1 score for SV calls made by $HQ3$, minimap2, and the Union model.}
\medskip
\begin{tabular}{|p{2.5cm}|p{5.0cm}|p{2.0cm}|p{1.8cm}|p{1.8cm}|p{1.8cm}|}
\hline
\bf{Dataset} & \bf{Truth set} & \bf{Method of alignment} & \bf{Precision} &\bf{Recall} & \bf{F1 score}\\
\hline
\multirow{3}{=}{HG002 reads to GRCh37}
& \multirow{3}{=}{GIAB Tier 1 calls}
  & minimap2 
  & $0.94$
  & $0.94$
  & $0.94$
  \\ \cline{3-6}
  &
  & $HQ3$ 
  & $0.94$
  & $0.94$
  & $0.94$
  \\ \cline{3-6}
  &
  & Union
  & $0.93$
  & $0.95$
  & $0.94$
  \\ \cline{3-6}
  \hline
\multirow{3}{=}{HG002 reads to CHM13}
& \multirow{3}{=}{\small comparing HG002 assembly to CHM13 (including centromere calls)}
  & minimap2 
  & $0.77$
  & $0.57$
  & $0.66$
  \\ \cline{3-6}
  &
  & $HQ3$ 
  & $0.75$
  & $0.58$
  & $0.65$
  \\ \cline{3-6}
  &
  & Union
  & $0.71$
  & $0.61$
  & $0.66$
  \\ \cline{3-6}
  \hline
\multirow{3}{=}{HG002 reads to CHM13}
& \multirow{3}{=}{\small comparing HG002 assembly to CHM13 (excluding centromere calls)}
  & minimap2 
  & $0.82$
  & $0.75$
  & $0.78$
  \\ \cline{3-6}
  &
  & $HQ3$ 
  & $0.82$
  & $0.74$
  & $0.78$
  \\ \cline{3-6}
  &
  & Union
  & $0.78$
  & $0.79$
  & $0.78$
  \\ \cline{3-6}
  \hline
\multirow{3}{=}{HG002 reads to GRCh37}
& \multirow{3}{=}{comparing HG002 assembly to GRCh37}
  & minimap2 
  & $0.78$
  & $0.76$
  & $0.77$
  \\ \cline{3-6}
  & 
  & $HQ3$ 
  & $0.79$
  & $0.74$
  & $0.77$
  \\ \cline{3-6}
  &
  & Union 
  & $0.74$
  & $0.79$
  & $0.77$
  \\ \cline{3-6}
  \hline
 \multirow{3}{=}{Simulated reads to CHM13}
& \multirow{3}{=}{Simulated SVs on chr 8 and X}
  & minimap2
  & $0.99$
  & $0.97$
  & $0.98$
  \\ \cline{3-6}
  &
  & $HQ3$ 
  & $0.99$
  & $0.97$
  & $0.98$
  \\ \cline{3-6}
  &
  & Union
  & $0.99$
  & $0.98$
  & $0.98$
  \\ \cline{3-6}
  \hline

\end{tabular}
\label{table:sv_precision_recall}
\end{table*}

\begin{figure*}[!h]
	\begin{minipage}{0.5\linewidth}
		\centering
		\includegraphics[width=\linewidth]{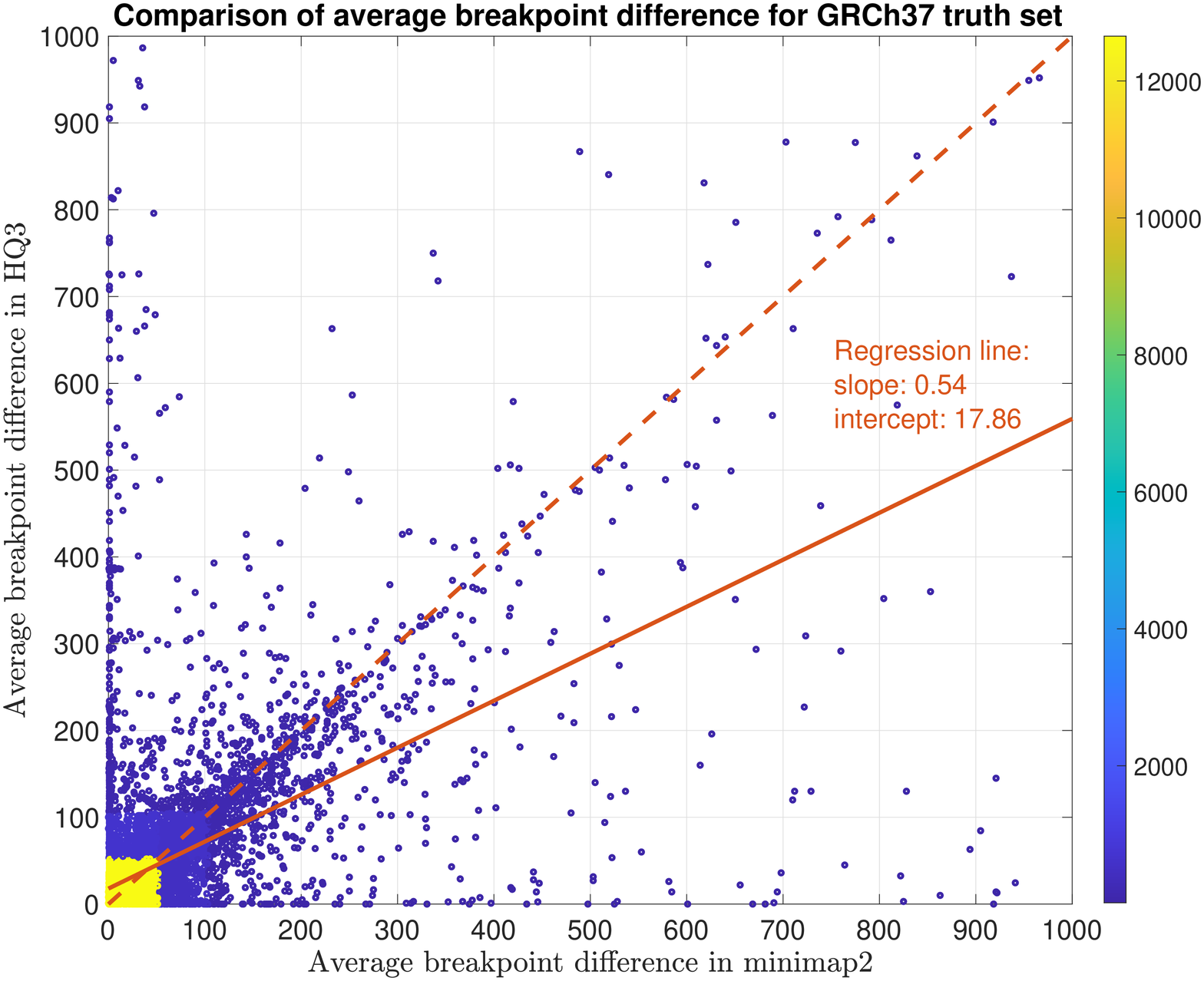}
		\label{fig:sv_hg002_breakpoints}
		(a)
	\end{minipage}\hfill
	\begin{minipage}{0.5\linewidth}
		\centering
		\includegraphics[width=\linewidth]{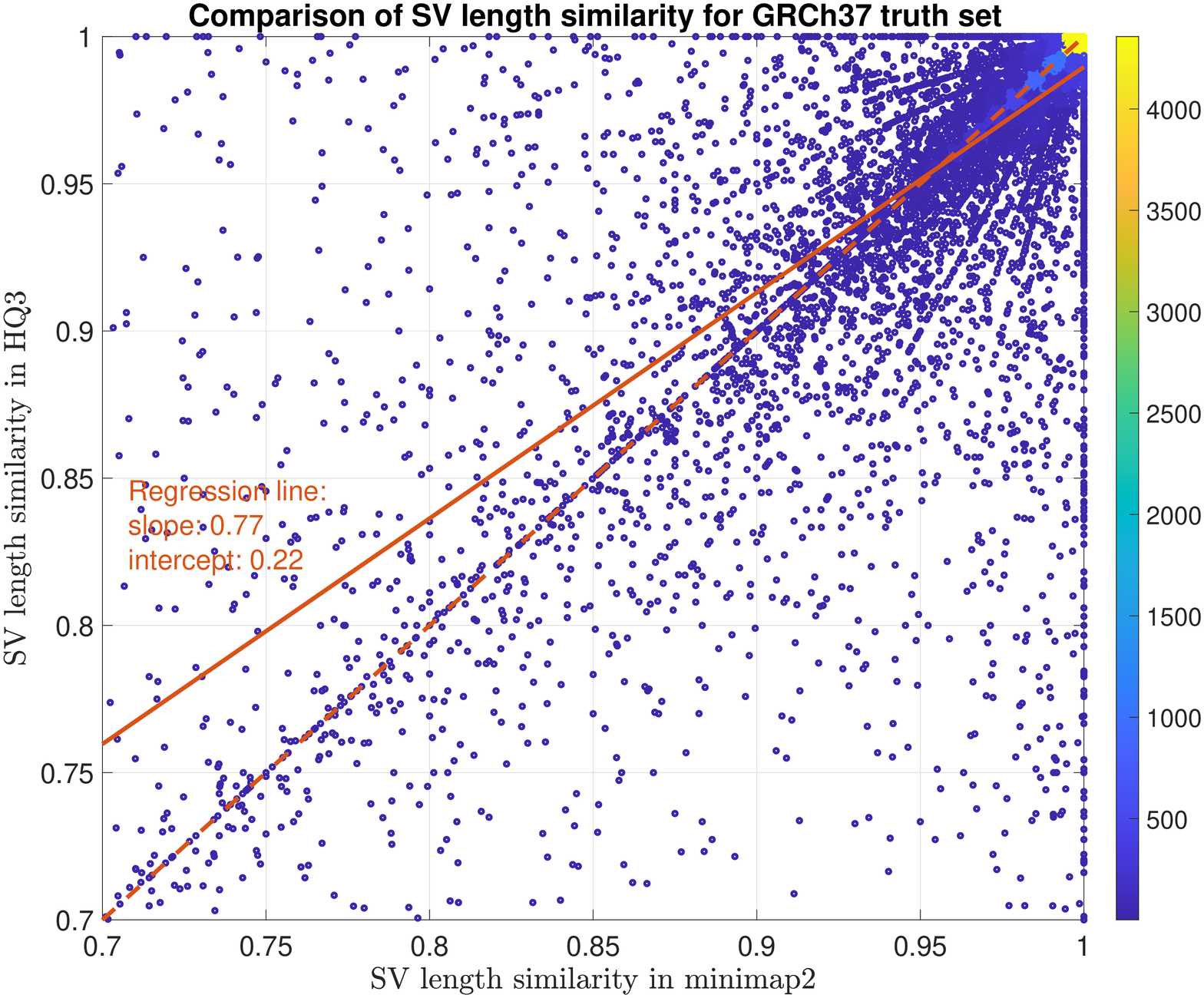}
		\label{fig:sv_hg002_svlength}
		(b)
	\end{minipage}\hfill
	\caption{{\bf SV quality comparison for common true positive calls in $HQ3$ and minimap2 against HG002-to-GRCh37 dipcall truth set.}
	(a) Comparison of SV breakpoint accuracy in $HQ3$ and minimap2 for common true positive calls. The difference of SV breakpoint is compared to the truth set generated from comparing HG002 haplotype-resolved assembly to GRCh37 build. A smaller difference represents better breakpoint accuracy. Therefore, slope of the regression line $<1$ represents better accuracy of $HQ3$ than minimap2 on average.
	(b) Comparison of SV length similarity in $HQ3$ and minimap2 for common true positive calls. The slope of the regression line $<1$ represents better SV length in minimap2 than $HQ3$ on average. However, this is due to a large density of SVs with length similarity $\geq 0.95$ in both minimap2 and $HQ3$. For length similarity less than $0.95$, $HQ3$ has better performance than minimap2.}
	\label{fig:sv_hg002}
\end{figure*}

\begin{figure*}[!h]
	\begin{minipage}{0.5\linewidth}
		\centering
		\includegraphics[width=\linewidth]{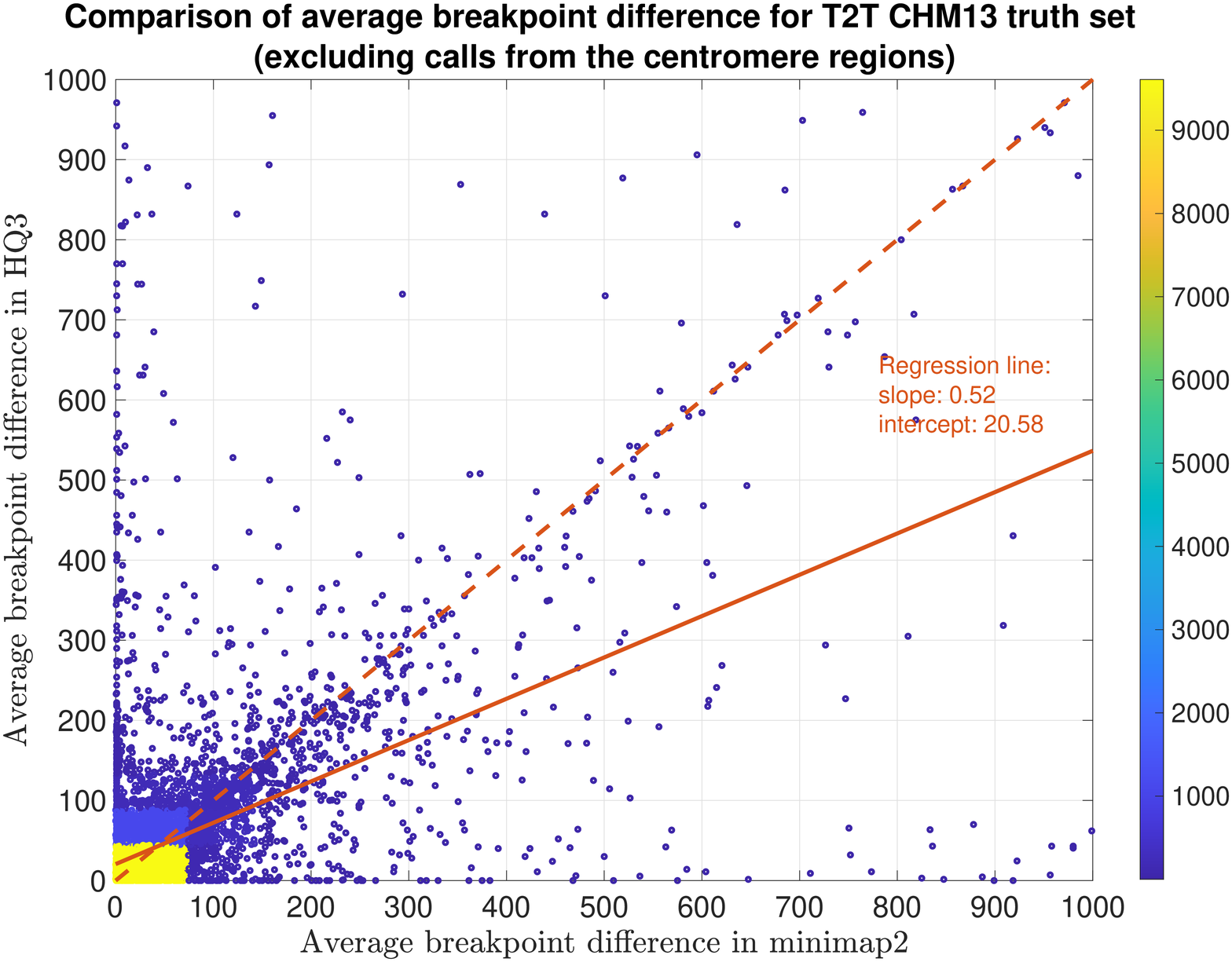}
		\label{fig:sv_chm13_breakpoints-centromere}
		(a)
	\end{minipage}\hfill
	\begin{minipage}{0.5\linewidth}
		\centering
		\includegraphics[width=\linewidth]{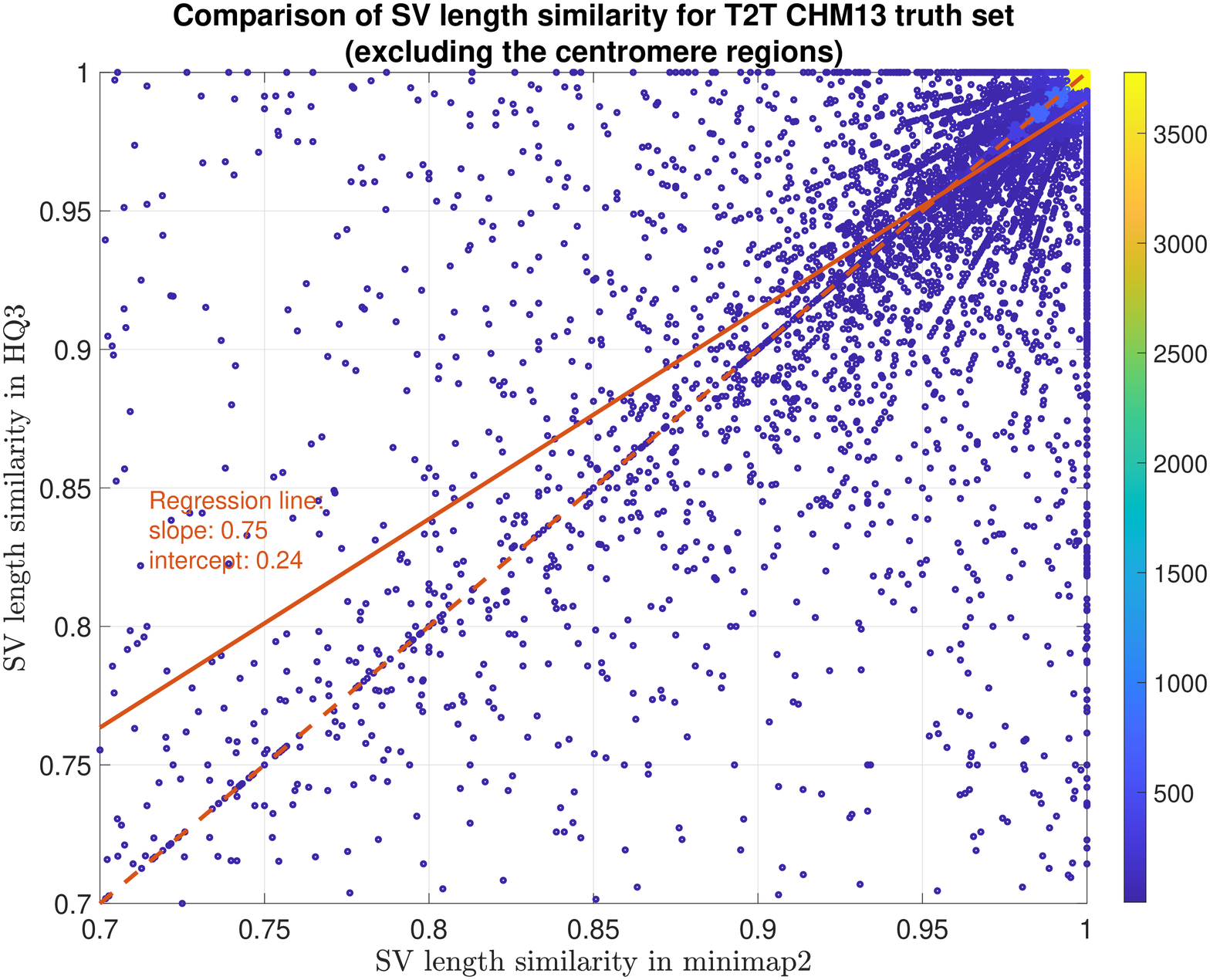}
		\label{fig:sv_chm13_svlength-centromere}
		(b)
	\end{minipage}\hfill
	\caption{{\bf SV quality comparison for common true positive calls in $HQ3$ and minimap2 against HG002-to-CHM13 dipcall truth set (excluding the centromere region).}
	(a) Comparison of SV breakpoint accuracy in $HQ3$ and minimap2 for common true positive calls. The difference of SV breakpoint is compared to the truth set generated from comparing HG002 haplotype-resolved assembly to T2T CHM13 build and the SV calls in the centromere region are excluded in this truth set. A smaller difference represents better breakpoint accuracy. Therefore, slope of the regression line $<1$ represents better accuracy of $HQ3$ than minimap2 on average.
	(b) Comparison of SV length similarity in $HQ3$ and minimap2 for common true positive calls. The slope of the regression line $<1$ represents better SV length in minimap2 than $HQ3$ on average. However, this is due to a large density of SVs with length similarity $\geq 0.95$ in both minimap2 and $HQ3$. For length similarity less than $0.95$, $HQ3$ has better performance than minimap2.}
	\label{fig:sv_chm13-centromere}
\end{figure*}

For SV calls from HG002 reads alignment to GRCh37 and benchmarking them against truth set generated by comparing HG002 haplotype-resolved assembly to GRCh37 build, minimap2 has precision $0.78$, recall $0.76$, and F1 score $0.77$ while $HQ3$ has precision $0.79$, recall $0.75$, and F1 score $0.77$. Out of $16462$ true positive calls in $HQ3$, $703$ ($4.27\%$) are made only in $HQ3$ with SV length similarity to the truth set greater than $0.7$ (default parameter in Truvari). However, $376/703$ calls that are captured by minimap2 with SV length similarity less than $0.7$ and $327/703$ calls that are uniquely made by $HQ3$.
Likewise, out of $16620$ true positive calls in minimap2, $861$ ($5.18\%$) are made only in minimap2 with SV length similarity greater than $0.7$. However, $524/861$ are captured by $HQ3$ with SV length similarity less than $0.7$ and $337/861$ are uniquely made by minimap2. A fine-grain analysis of the common true positive calls by minimap2 and $HQ3$ in Figure \ref{fig:sv_hg002}a, shows that a major density of SV calls ($81.85\%$) have difference in breakpoint below $50$ in both minimap2 and $HQ3$, and minimap2 has marginally better performance in terms of lower difference in breakpoint of SVs for difference in breakpoint below $50$. Whereas, for a large difference in the SV breakpoint (greater than $50$), $HQ3$ is better in terms of the breakpoint accuracy of the SV calls (on average across all SV calls). Therefore, $HQ3$ improves the SV breakpoint for the rest $18.15\%$ calls that have high difference in breakpoints.
Further, Figure \ref{fig:sv_hg002}b demonstrate that $HQ3$ has better SV length similarity when the length similarity is below $0.95$ which corresponds to $21.82\%$ calls.

\section{Discussion}

HQAlign method is an alignment method designed for the detection of structural variants for nanopore sequencing reads. HQAlign provides alignment that outperforms the recent minimap2 aligner in terms of the accuracy and quality of the alignments. The SV calling from HQAlign is also at par with minimap2 in terms of F1 score and it outperforms minimap2 SV calls in terms of the quality of SVs measured in breakpoint accuracy and SV length similarity. Moreover, there are many complementary SVs captured by HQAlign that are missed by minimap2 alignments.

The reason for this improvement in the performance of alignment and SV calling with HQAlign is that it takes into account the underlying physics of nanopore sequencer through the $Q$-mer map, which could be one of the major causes of the high error rates in nanopore sequencing, and also it focuses on a narrow region of the genome (where the read aligns in nucleotide domain) for alignment with quantized sequences. Further, this pipeline is adapted specifically for the detection of SVs.
We demonstrated how HQAlign utilizes the bias of $Q$-mer map without accessing the raw current signal of nanopore sequencer by translating the basecalled nucleotide sequences to quantized current level (of finite alphabet size) sequences. This improvement help in detecting several SVs that are missed by minimap2 due to high error rates in the nanopore reads. Further, the recall rate for SV detection can be improved by combining the complementary calls from both $HQ3$ and minimap2 in the union model at the same F1 score.

\section*{Competing interests}
  The authors declare that they have no competing interests.

\section*{Author's contributions}
DJ,  SK, and SD conceived the original idea and developed the project.
DJ led the development of the software tool and its open-source development.
MC helped with SV metrics, datasets, and SV comparison analysis between methods.
DJ performed the analysis on the various datasets for both alignment and SV calling. All the authors wrote the manuscript.

\section*{Acknowledgements}
SD and DJ were supported in part by National Science Foundation grant 1705077.
MC was supported by National Institutes of Health grant R01HG011649.
SK was supported in part by National Institutes of Health grant 1R01HG008164 and National Science Foundation grants 1651236 and 1703403.

\bibliographystyle{naturemag}
\bibliography{mybib.bib}

\appendix

\newpage
\section{Supplementary material}

\subsection{Quantization method from QAlign \cite{joshi2021qalign}}

The nucleotide sequences are inferred from the nanopore current signals by basecallers, therefore, using a $Q$-mer map to translate the basecalled sequences to the current levels implicitly maintains all of the ``equivalent'' basecalled sequences that could be inferred from the observed current levels. These current levels can be quantized to an alphabet of finite size.

\begin{figure*}[!h]
	\centering
	\includegraphics[width=\linewidth]{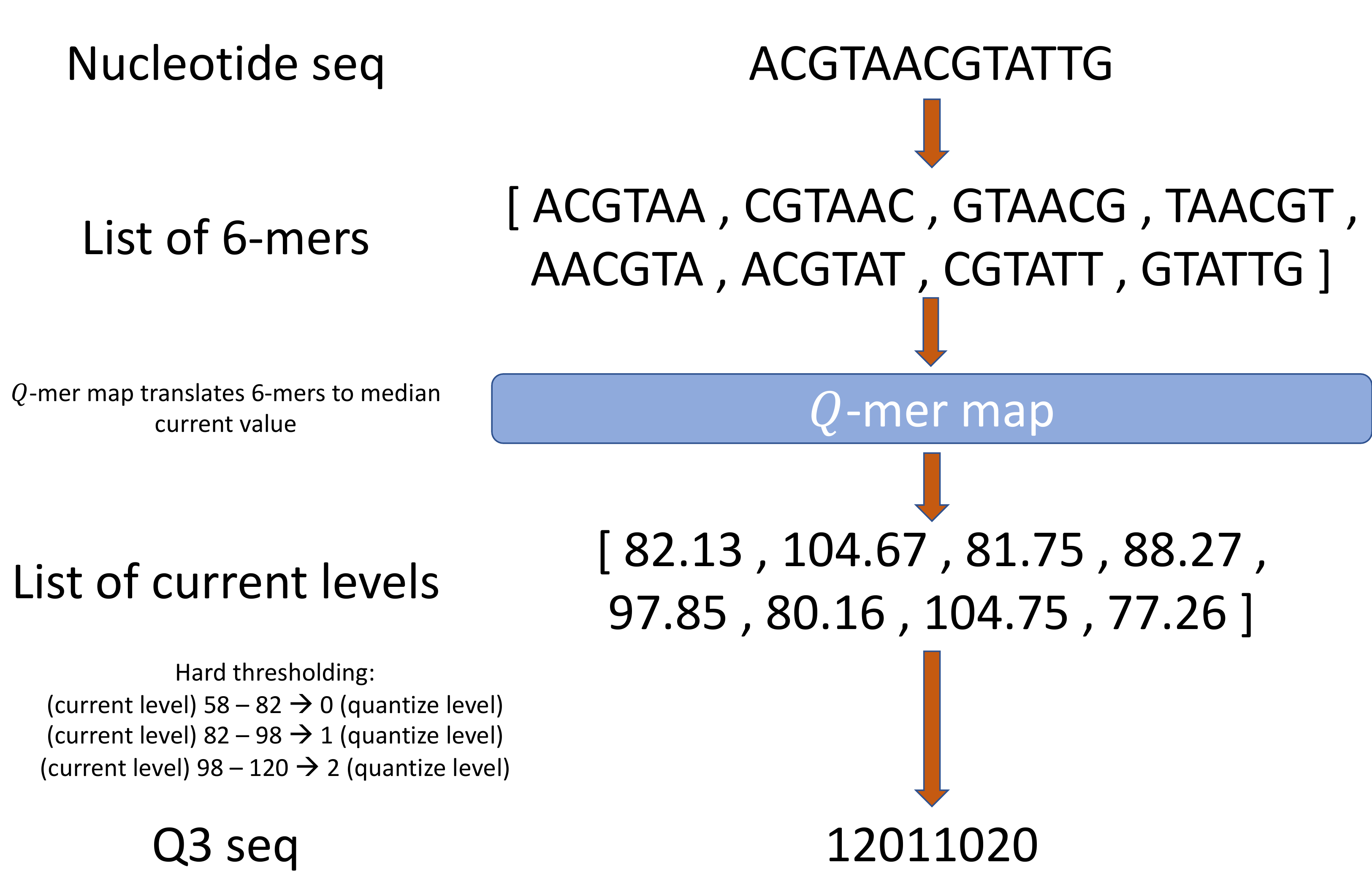}
	\caption{\textbf{An example for the Quantization method for QAlign.}
	The nucleotide sequences are first translated to current level sequences using the $Q$-mer map, and then the (continuous) current level sequences are quantized to finite levels (\emph{e.g.} three levels for $HQ3$) by hard thresholding the current levels.
	}
	\label{fig:quantization}
\end{figure*}

Mathematically, the quantization process is as follows.
Let $\Sigma = \{A,C,G,T\}$ be the alphabet of nucleotide sequences.
For a symbol $s\in\Sigma$, let $\bar s$ be the Watson-Crick complement of $s$.
A string $x=s_1 s_2 \dots s_n$ over $\Sigma$ is called a nucleotide sequence, where $|x|=n$ is the string length and the reverse complement of $x$ is $\overline x = \overline{s_1s_2\dots s_n} = \overline s_n \overline s_{n-1} \dots \overline s_1$.
Let $p(x)$ be a list of all $Q$-mers (e.g. $Q{=}6$) in the string $x$, sorted by their occurrences. For example, $p(x) = k_1k_2\dots k_{n-Q+1}$ and each $Q$-mer $k_i = s_is_{i+1}\dots s_{i+Q-1}$ for $i=1,2,\dots,n-Q+1$. Now, we define $f: \Sigma^Q \rightarrow \mathbb R$ as the $Q$-mer map \footnote{$Q$-mer map is determined by the chemistry of the nanopore flow cell, and is therefore dataset dependent, \emph{i.e.}, the $Q$-mer map for sequencing using R9 flow cell is different from $Q$-mer map for sequencing using R9.4.1 flow cell. The $Q$-mer maps used in this work are generated by Nanopolish ({https://github.com/jts/nanopolish}).}, which is a deterministic function that translates each $Q$-mer $(k_i)$ to the (median) current level (Figure 1b).
Now, let $C(x)=c_1c_2\dots c_{n-Q+1}$ be the sequence of the current levels, so that $c_i = f(k_i)$ for $i=1,2,\dots,n-Q+1$.
The current sequence $C(x)$ can be further quantized into $w(x)=q_1q_2 \dots q_{n-Q+1}$ by applying hard thresholding function $q_i = g(c_i)$. The thresholding can be ternary ($q_i \in \{0, 1, 2\}$) for $HQ3$ (Figure 1c and Supplemental Figure \ref{fig:quantization}).
We define $w(\overline x)$ as the quantized reverse complementary of sequence $x$, so $\overline w(x)=w(\overline x)$. Supplemental Figure \ref{fig:quantization} explains this process using a toy example.

\subsection{Generalization of HQAlign method}

\subsubsection{Initial alignment}
The nucleotide query $x$ is aligned to a set of nucleotide target sequences $t = (t_1,t_2,\dots,t_m)$ using Minimap2. This is similar to aligning a read to a genome which has several chromosome sequences. This step identifies the region of interests on the target $t$, say, $t_j[s_i:e_i]$, where $t_j$, $j \in \{1,2,\dots,m\}$ represent alignment to one or more target chromosomes that $x$ aligns to, $i \in \{1,2,3,\dots\}$ represent represent one or more alignments to chromosome $j$, $s_i$ and $e_i$ are the corresponding start and end location of each alignment $i$ on the target $t_j$, respectively.

\subsubsection{Hybrid alignment}
In this step, the query $x$ is re-aligned to an extended region of interest on the target $t_j[s_i^q:e_i^q]$ using the QAlign method, where $s_i^q = s_i - b_i$ and $e_i^q = e_i + b_i$, $b_i = (1-f_i+0.25)n$ is an appended extension of the region of interest on target, $f_i = (e_i-s_i)/n$ is the fraction of read aligned in initial step, and $n$ is the length of the query $x$. The nucleotide query $x$ and the nucleotide extended target $t_j[s_i^q:e_i^q]$ are converted to the quantized query $x^q$, quantized reverse complement query $\overline x^q$ and quantized extended target $t_j^q[s_i^q:e_i^q]$, respectively, using the quantization method demonstrated in QAlign. These quantized sequences are then aligned using modified minimap2 pipeline.

\newpage

\begin{table*}[!h]
\subsection{Accessing HQAlign on github}
HQAlign requires python 3, and the installation guideline can be found on github.\\
The software is available at: \url{https://github.com/joshidhaivat/HQAlign.git}\\\\

\texttt{usage: python hqalign.py [-h] -r REF -i READS -o OUTPUT [-t THREADS] [-k KMER]}\\\\
arguments:\\
\begin{tabular}{p{6.5cm}p{10cm}}
\texttt{-h, --help} & show this help message and exit\\
\texttt{-r REF, --ref REF} &reference genome filename in fasta format\\
\texttt{-i READS, --reads READS} &directory location of read files in fasta format (with file extension \texttt{.fasta})\\
\texttt{-o OUTPUT, --output OUTPUT} &location of directory of output files\\
\texttt{-t THREADS, --threads THREADS} &maximum number of parallel threads (default=4)\\
\texttt{-k KMER, --kmer KMER} &minimizer length for hybrid step (default=18)
\end{tabular}
\end{table*}

\begin{figure*}[!h]
	\centering
	\includegraphics[width=\linewidth]{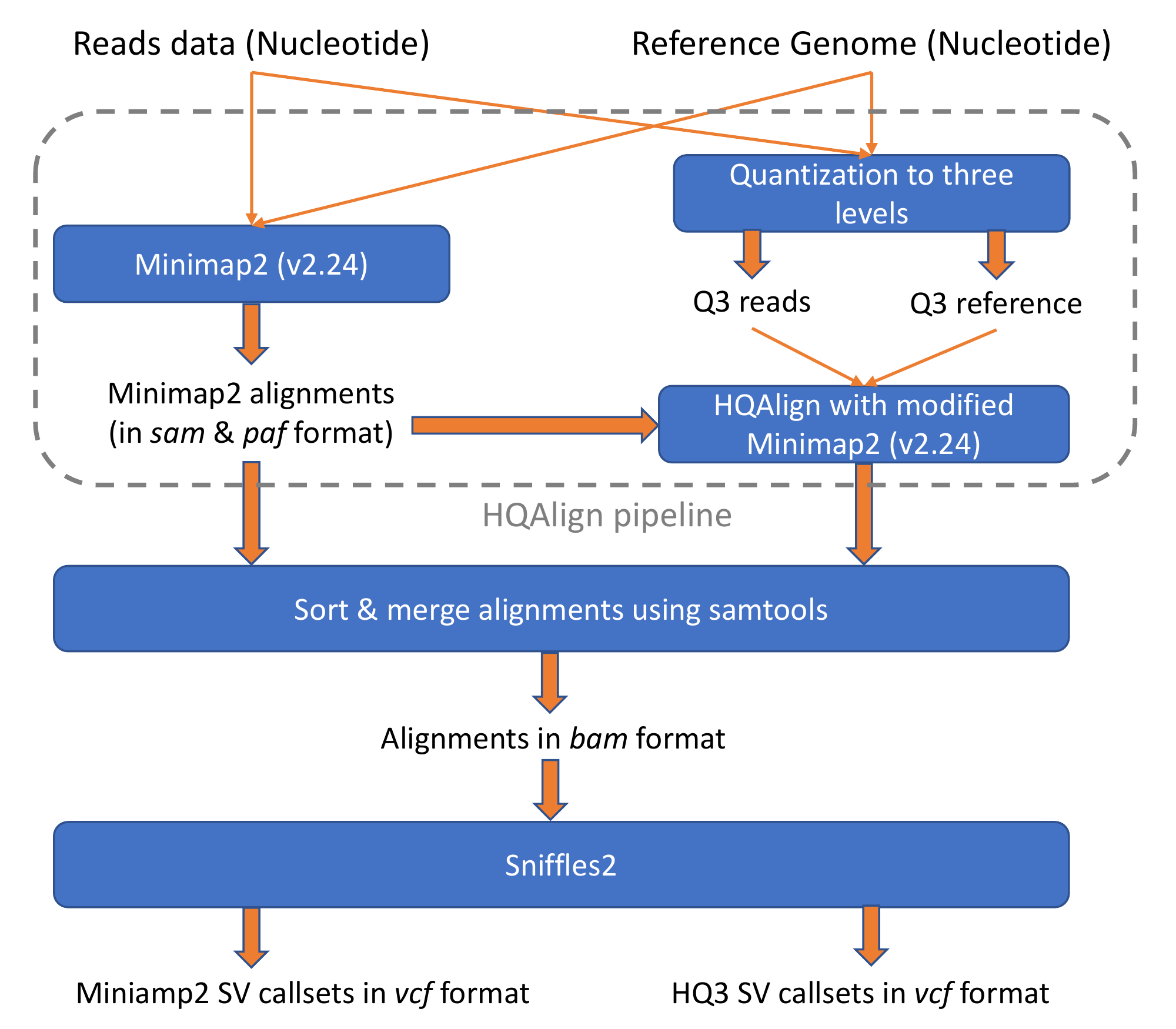}
	\caption{Complete pipeline for SV calling using minimap2 and HQAlign.
	}
	\label{fig:pipeline_flowchart}
\end{figure*}

\newpage

\newpage

\begin{figure*}[!h]
	\begin{minipage}{0.49\linewidth}
		\centering
		\includegraphics[width=\linewidth]{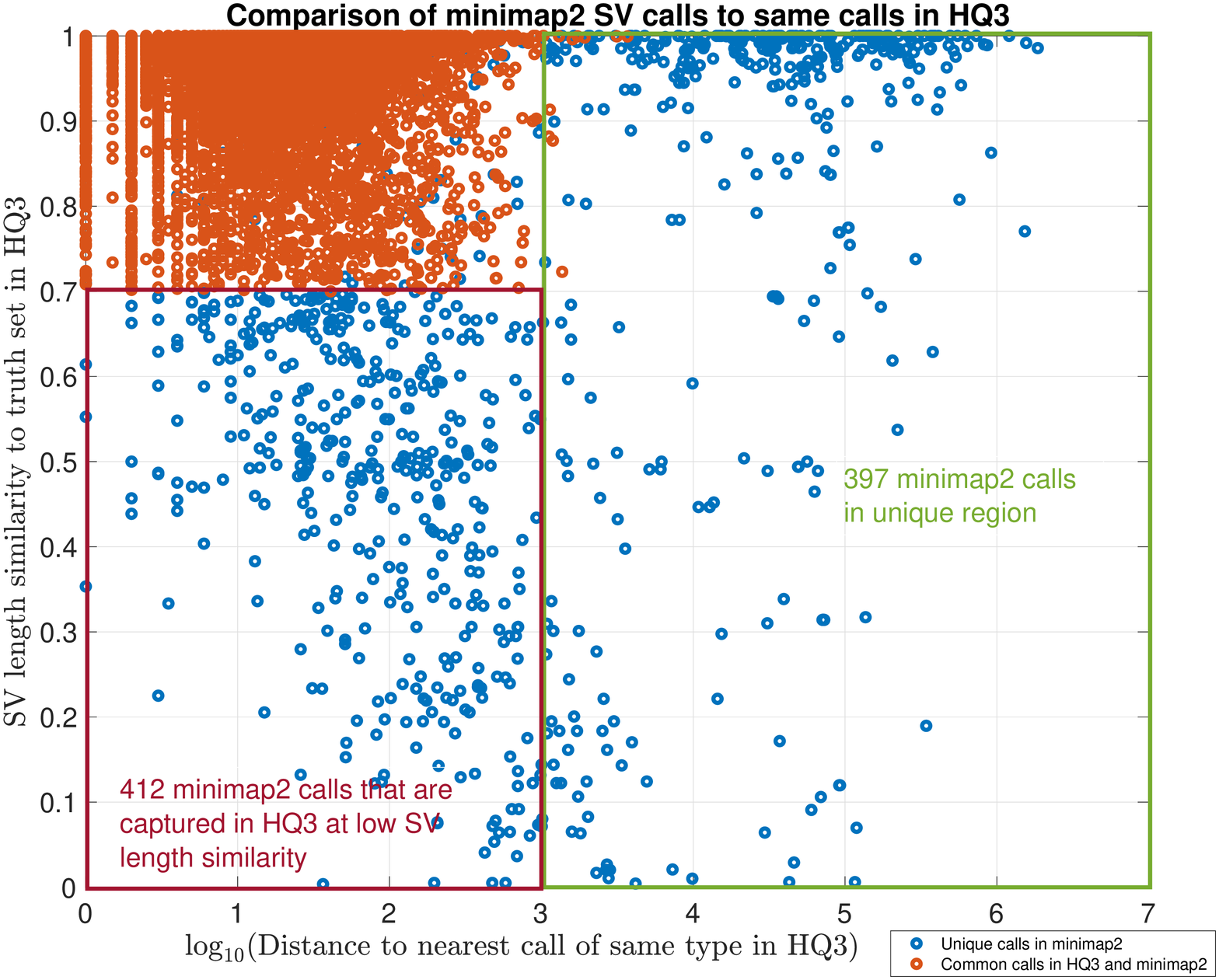}
		\label{fig:chm13_acgt_unique_calls}
		(a)
	\end{minipage}\hfill
	\begin{minipage}{0.49\linewidth}
		\centering
		\includegraphics[width=\linewidth]{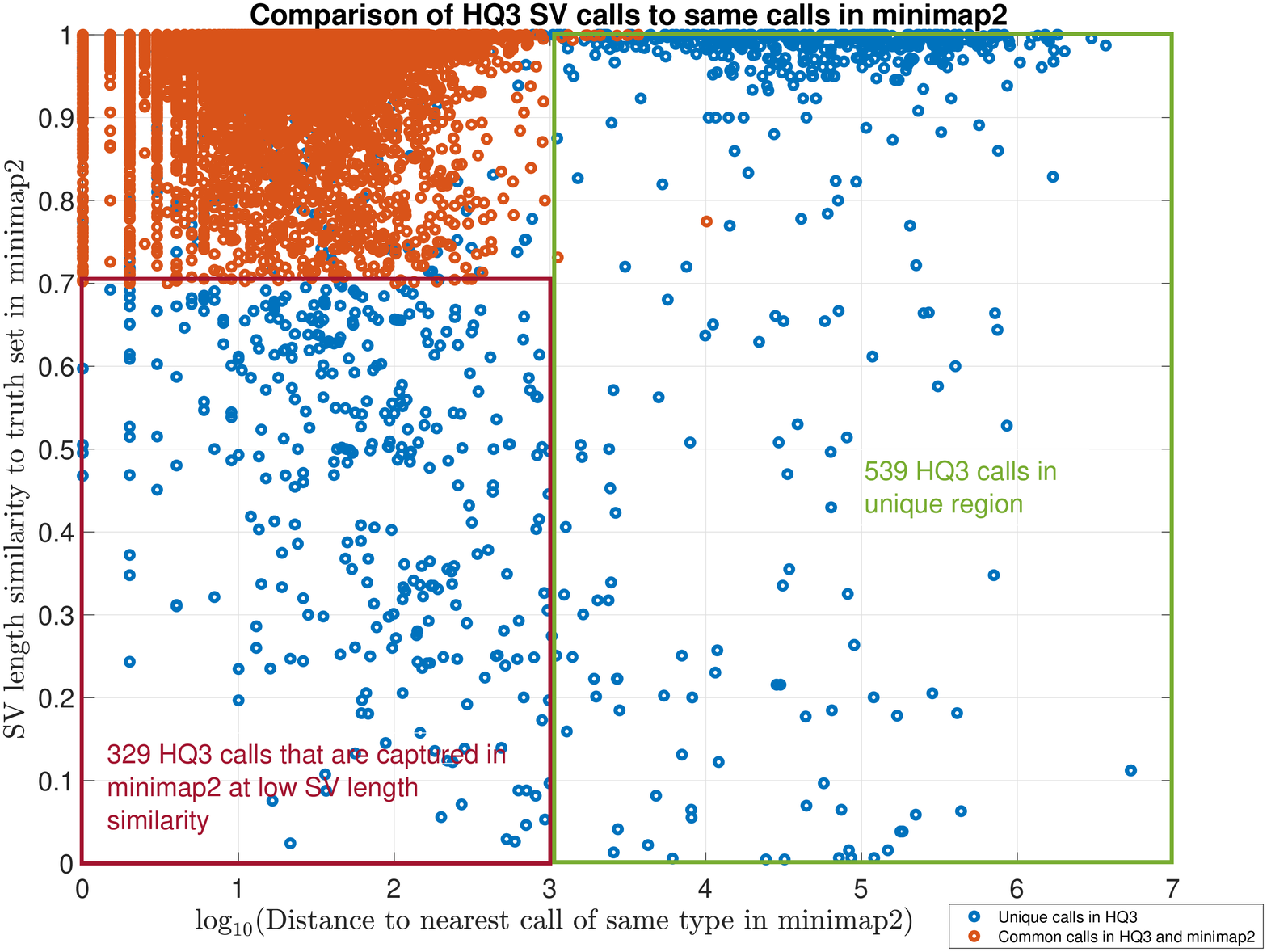}
		\label{fig:chm13_hq3_unique_calls}
		(b)
	\end{minipage}\hfill
	\caption{{\bf Comparison of SV calls made by minimap2 and $HQ3$ to other method.}
	(a) For the complementary calls (in blue) and common calls (in red) made by minimap2, this figure compares SV length similarity and distance to nearest SV in HQ3 of the same type. $397$ complementary calls made by minimap2 are in unique region, whereas $412$ complementary calls in minimap2 are captured in neighboring region (within 1000 bp) in $HQ3$ but with a low SV length similarity.
	(b) For the complementary calls (in blue) and common calls (in red) made by $HQ3$, this figure compares SV length similarity and distance to nearest SV in minimap2 of the same type. $539$ complementary calls made by $HQ3$ are in unique region, whereas $329$ complementary calls in $HQ3$ are captured in neighboring region (within 1000 bp) in minimap2 but with a low SV length similarity.}
	\label{fig:chm13_uniquecalls}
\end{figure*}

\begin{figure*}[!h]
	\begin{minipage}{0.49\linewidth}
		\centering
		\includegraphics[width=\linewidth]{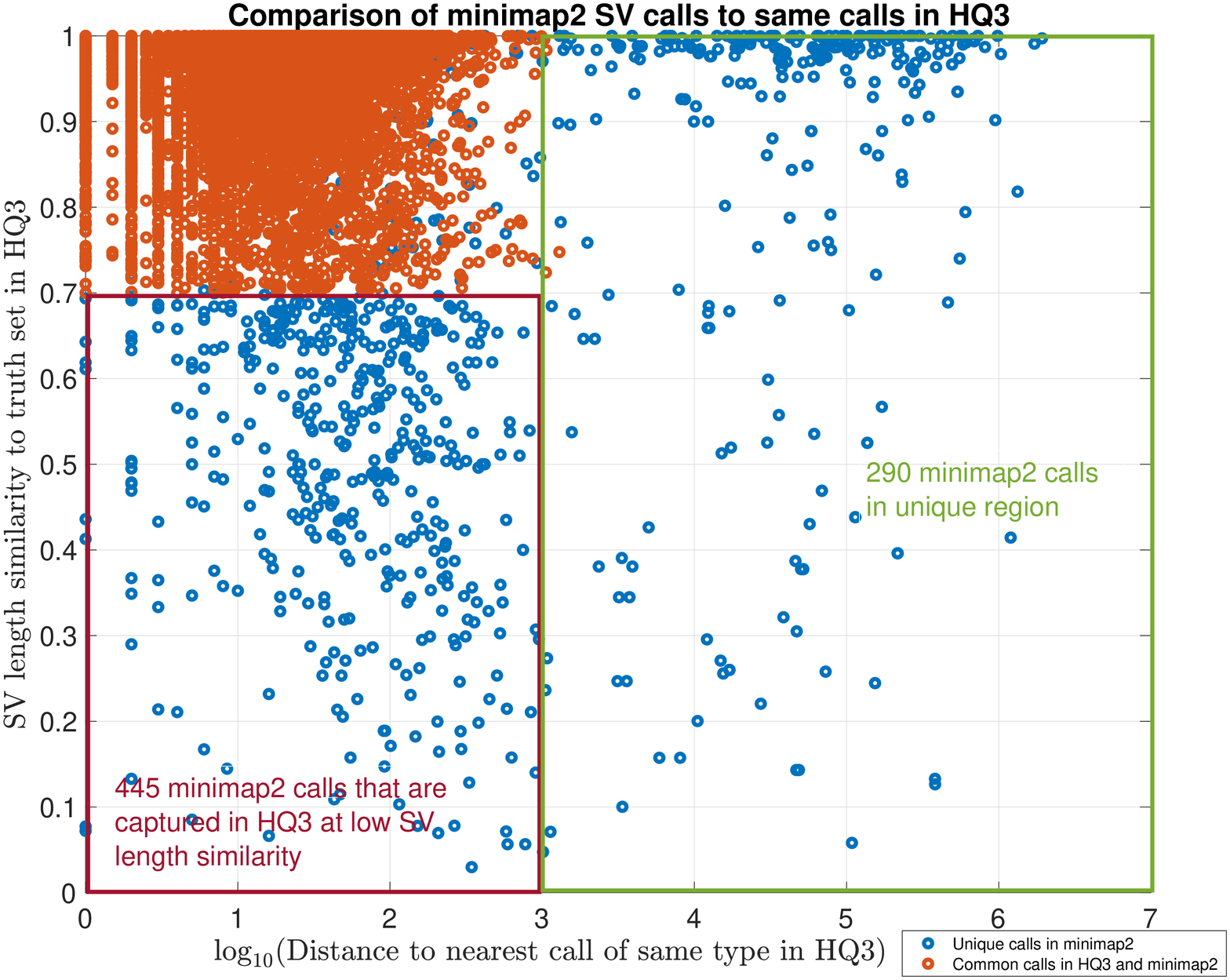}
		\label{fig:hg002_acgt_unique_calls}
		(a)
	\end{minipage}\hfill
	\begin{minipage}{0.49\linewidth}
		\centering
		\includegraphics[width=\linewidth]{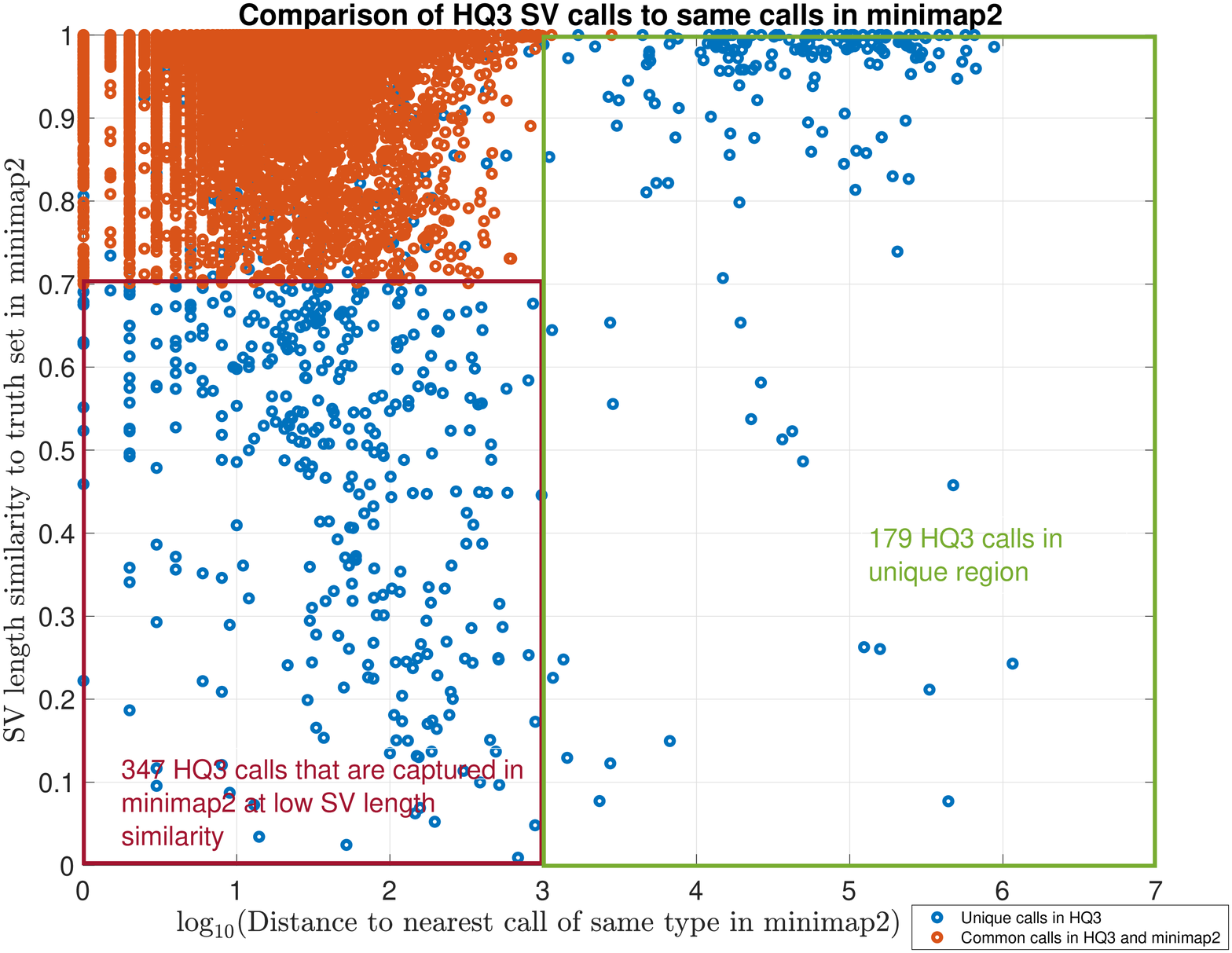}
		\label{fig:hg002_hq3_unique_calls}
		(b)
	\end{minipage}\hfill
	\caption{{\bf Comparison of SV calls to HG002 truth set.}
	(a) For the complementary calls (in blue) and common calls (in red) made by minimap2, this figure compares SV length similarity and distance to nearest SV in HQ3 of the same type. $290$ complementary calls made by minimap2 are in unique region, whereas $445$ complementary calls in minimap2 are captured in neighboring region (within 1000 bp) in $HQ3$ but with a low SV length similarity.
	(b) For the complementary calls (in blue) and common calls (in red) made by $HQ3$, this figure compares SV length similarity and distance to nearest SV in minimap2 of the same type. $179$ complementary calls made by $HQ3$ are in unique region, whereas $347$ complementary calls in $HQ3$ are captured in neighboring region (within 1000 bp) in minimap2 but with a low SV length similarity.}
	\label{fig:hg002_uniquecalls}
\end{figure*}


\end{document}